%% file: paper.tex
\documentclass[letterpaper,twocolumn,10pt]{article}
\usepackage{usenix-2020-09}

\usepackage{tikz}
\usepackage{amsmath}
\usepackage{xspace}
\usepackage{cite}
\usepackage{amsmath,amssymb,amsfonts}
\usepackage{graphicx}
\usepackage{textcomp}
\usepackage{xcolor}
\def\BibTeX{{\rm B\kern-.05em{\sc i\kern-.025em b}\kern-.08em
		T\kern-.1667em\lower.7ex\hbox{E}\kern-.125emX}}
\usepackage{tikz}
\usepackage{amsmath}
\usepackage{filecontents}
\usepackage{tikz}
\usepackage{amsmath}
\usepackage{filecontents}
\usepackage{balance} 
\usepackage{color}
\usepackage{amsmath, amssymb}
\usepackage{makecell}
\usepackage{multirow}
\usepackage{cite}
\usepackage{amsmath,amssymb,amsfonts}
\usepackage{graphicx}
\usepackage{textcomp}
\usepackage{xcolor}
\usepackage{times}
\usepackage{listings}
\usepackage{graphicx}
\usepackage{subfig}
\usepackage{booktabs}
\usepackage{url}
\usepackage{mwe}
\usepackage{algorithm,algpseudocode}
\usepackage{enumitem}

\newcommand{\sysname}{{HiStore}\xspace}

\begin{document}

\title{\sysname: Rethinking Hybrid Index in RDMA-based Key-Value Store}

\author{\rm Paper ID: 157}

\author{
{\rm Shukai Han, Mi Zhang, Dejun Jiang, Jin Xiong}\\
SKL Computer Architecture, ICT, CAS; University of Chinese Academy of Sciences\\
\{hanshukai, zhangmi, jiangdejun, xiongjin\}@ict.ac.cn
} 

\maketitle

\input{0_abstract}
\input{1_introduction}
\input{2_motivation}
\input{3_design}
\input{4_implement}
\input{5_evaluation}
\input{6_relatedwork}
\input{7_conclusion}

\bibliographystyle{abbrv}
\bibliography{reference}

\end{document}

%% file: 0_abstract.tex
\begin{abstract}
RDMA (Remote Direct Memory Access) is widely exploited in building key-value stores to achieve ultra
    low latency. In RDMA-based key-value stores, the indexing time takes a large fraction (up
    to 74\%) of the overall operation latency as RDMA enables fast data accesses. However, the
    single index structure used in existing RDMA-based key-value stores, either hash-based or sorted
    index, fails to support range queries efficiently while achieving high performance for single-point operations.
    In this paper, we reconsider the
    adoption of hybrid index in the key-value stores based on RDMA, to combine the benefits of hash
    table and sorted index. We propose \sysname, an RDMA-based key-value store using hash table for
    single-point lookups and leveraging skiplist for range queries. To maintain strong
    consistency in a lightweight and efficient approach, \sysname introduces index groups where a
    skiplist corresponds to a hash table, and asynchronously applies updates to the skiplist within
    a group. Guided by previous work on using RDMA for key-value services, \sysname dedicatedly
    chooses different RDMA primitives to optimize the read and write performance.
    Furthermore, \sysname tolerates the failures of servers that maintain index structures with index
    replication for high availability. Our evaluation results demonstrate that \sysname improves the
    performance of both GET and SCAN operations (by up to 2.03x) with hybrid index.
\end{abstract}

%% file: 1_introduction.tex
\section{Introduction}

Key-value store is a vital component in modern data centers for building various applications.  Many
existing systems, such as databases, social networks, online retail, and web services, use
key-value stores as the storage engines~\cite{chang06bigtable, atikoglu2012workload,
nishtala2013scaling, lai15atlas, decandia17dynamo, caofast20}. The simple interfaces (e.g., PUT,
GET, SCAN) and high performance of key-value stores enable users to efficiently store and access a
large volume of data.

Remote Direct Memory Access (RDMA) is widely studied to improve the performance of key-value stores
in recent years, namely \textit{RDMA-based} or \textit{RDMA-enabled} key-value
stores~\cite{mitchell13pilaf, kalia14herd, wei15drtm, dragojevic14farm, zamanian2017nam-db,
li2017kv-direct}.
RDMA communication provides two types of primitives, \textit{one-sided} verbs allow to directly access
data in remote memory without involving the server CPU, and \textit{two-sided} verbs
enable fast message-based data transfer (like the conventional network protocols).
The RDMA-based key-value stores utilize different RDMA verbs to provide key-value services, which
can be classified into \textit{server-centric}, \textit{client-direct}, and \textit{hybrid-access}
designs. The server-centric stores only use two-sided verbs to support key-value operations, while
the client-direct designs only leverage one-sided RDMA reads and writes. The hybrid-access stores
exploit both one-sided and two-sided verbs for CPU efficiency and low latency, which combines the
benefits of the server-centric and client-direct designs.
 
When handling key-value operations, indexing plays a critical role in the whole process, especially
for RDMA-based systems. As RDMA enables fast data access, the indexing performance largely
determines the overall performance of single-point operations. Our analysis demonstrates that the indexing latency
accounts for 49-74\% of the total operation time (for PUT and GET) in an RDMA-based
key-value store. Therefore, it is important to reduce indexing latency when building low-latency
key-value stores based on RDMA.

Existing RDMA-based key-value stores typically leverage single index for key lookups, either
hash-based~\cite{mitchell13pilaf, dragojevic14farm, kalia14herd, wang15hydradb, wei15drtm,
kalia2016fasst, li2017kv-direct, cassell17nessie, zuo21race} or sorted index (e.g., tree-backed,
skiplist)~\cite{dragojevic15farm, mitchell16cell, chen16drtmr, kalia19erpc, ziegler19, wei20xstore}.
The hashing index locates a key-value pair based on the hash value of the key, which provides fast
single-key lookups and can be easily completed using one-sided verbs.
For example, RACE~\cite{zuo21race} hashing index executes all index requests using only one-sided
RDMA verbs.
However, it is hard to deal with range queries (i.e., SCAN) based on hashing index. 
Thus, some RDMA-based key-value stores choose sorted index which maintains the order of key-value
pairs to provide efficient range queries. Though sorted index supports rich key-value operations,
sorted index incurs longer latency than hashing index for single-key lookups.
Searching along a sorted index requires involvement of the server CPU to complete within one round
trip; otherwise, it incurs multiple round trips if only using one-sided RDMA reads.
Our study shows that when there is no specific optimization, the indexing latency of sorted index
can be as high as three times of the latency using a hash table.

To provide rich key-value services with low latency, it is promising to combine a hash table and
a sorted index for single-key lookups and range queries respectively in RDMA-based key-value stores.
HiKV~\cite{xia17hikv} realizes the idea of hybrid index on a single server with hybrid memory and
demonstrates the performance improvement of key-value operations.
However, adopting hybrid index in RDMA-based key-value stores poses new challenges.
First, the index management requires to keep different index structures consistent with the
key-value items in an efficient manner. Meanwhile, if the storage system replicates the index for
high availability, the replicas should be updated consistently.
Thus, how to mitigate the management overhead of hybrid index with strong consistency remains an
open issue. Furthermore, as failures in distributed systems are commonplace, the key-value stores
using hybrid index should consider how to support key-value services and reconstruct the index in
the present of failures.  

We present \sysname, an RDMA-based key-value store which combines a hash table and a sorted index
(i.e., skiplist) to support rich key-value operations and achieve high indexing performance.
\sysname combines different indexes in one \textit{index group} (a unit of hybrid index) to
efficiently manage them with strong consistency.
It allows hybrid-access using different RDMA primitives for CPU efficiency and low latency.
That is, \sysname uses two-sided verbs for write operations to guarantee strong consistency; for
read requests, it uses one-sided verbs for GET operations to bypass the server CPU, while leveraging
two-sided verbs for SCAN operations to reduce the number of round trips. 
To minimize the write latency, \sysname batches the index updates and applies the updates to the
skiplist asynchronously.
Moreover, \sysname achieves high availability with index replication to protect index structures
against failures.
To the best of our knowledge, \sysname is the fist RDMA-based key-value store that leverages
hybrid index for key-value services.

We summarize our contributions as follows.
\begin{itemize}[leftmargin=*]
    \item We analyze the usage of single index in RDMA-based key-value store, and
        motivate the adoption of hybrid index by comparing the performance of different indexes.
    \item We propose to combine a hashing index and a sorted index in one index
        group for efficient index management. Each index group consistently updates the indexes
        within the group, and reduces the write latency with asynchronously updating the skiplist in a batch. 
    \item We add a replica of skiplist to each group for high availability. The index group
        can efficiently support rich key-value services in the face of single-server failures.
    \item We implement \sysname using eRPC~\cite{kalia19erpc} in two-sided communications and
        conduct extensive experiments to evaluate its read and write performance. 
\end{itemize}

We will make the source code of \sysname publicly available after this paper is accepted.



%% file: 2_motivation.tex
\section{Background and Motivation}
\label{sec:motivation}

\begin{figure*}[t]
	\centering
	\subfloat[Server-centric design]{
		\label{fig:server-centric}
		\includegraphics[width=0.31\textwidth]{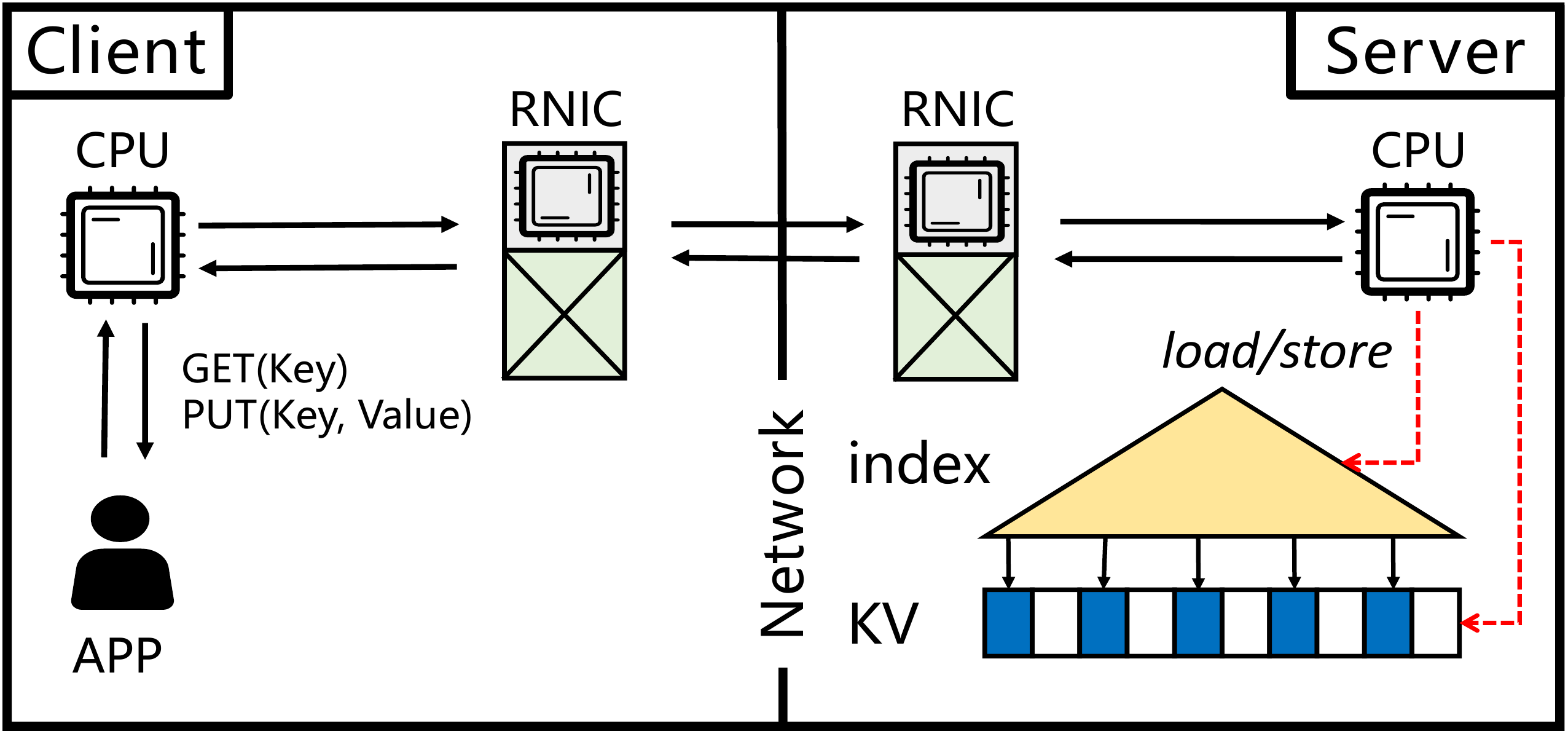}}
	\subfloat[Client-direct design]{
		\label{fig:client-direct}
		\includegraphics[width=0.31\textwidth]{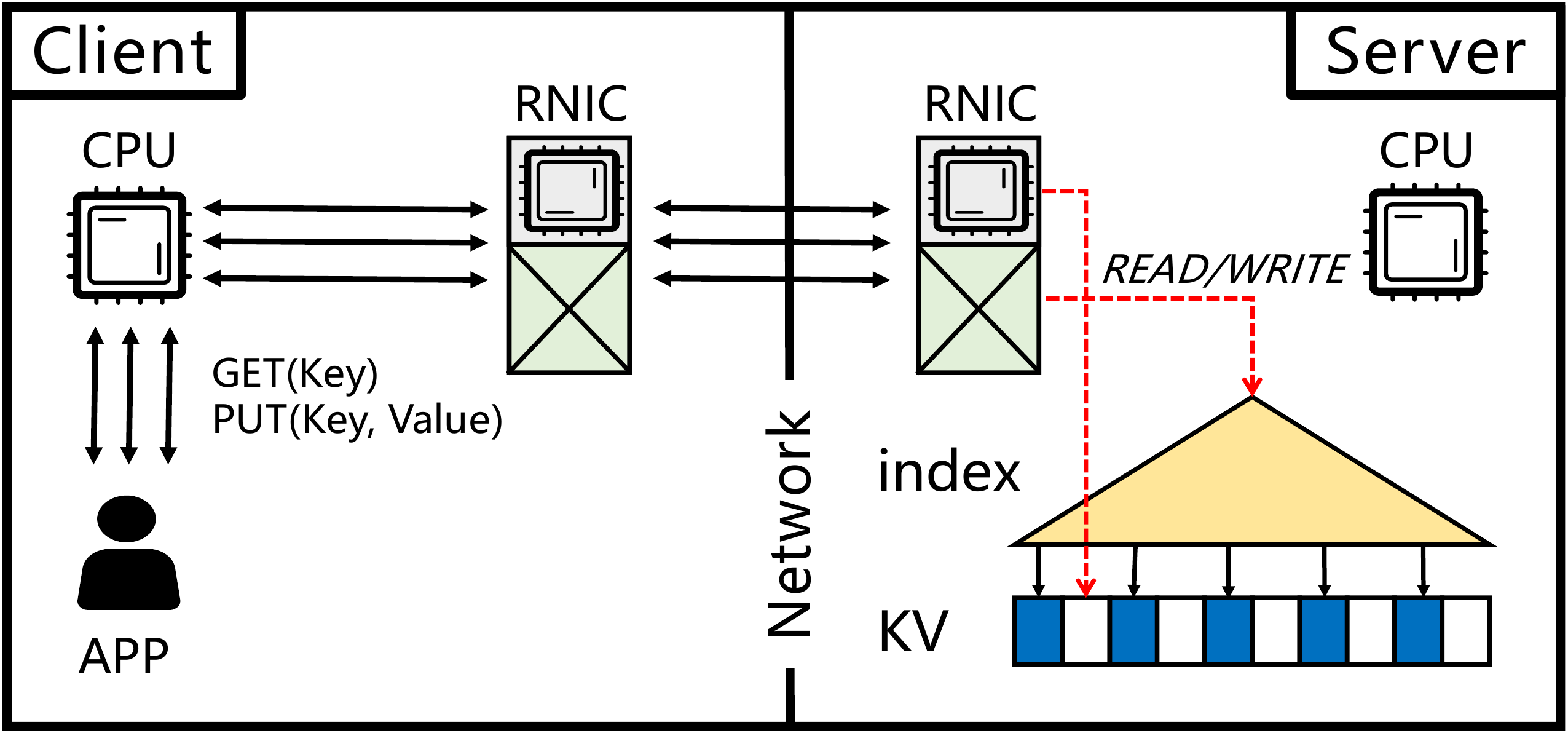}}
	\subfloat[Hybrid-access design]{
		\label{fig:hybrid-access}
		\includegraphics[width=0.31\textwidth]{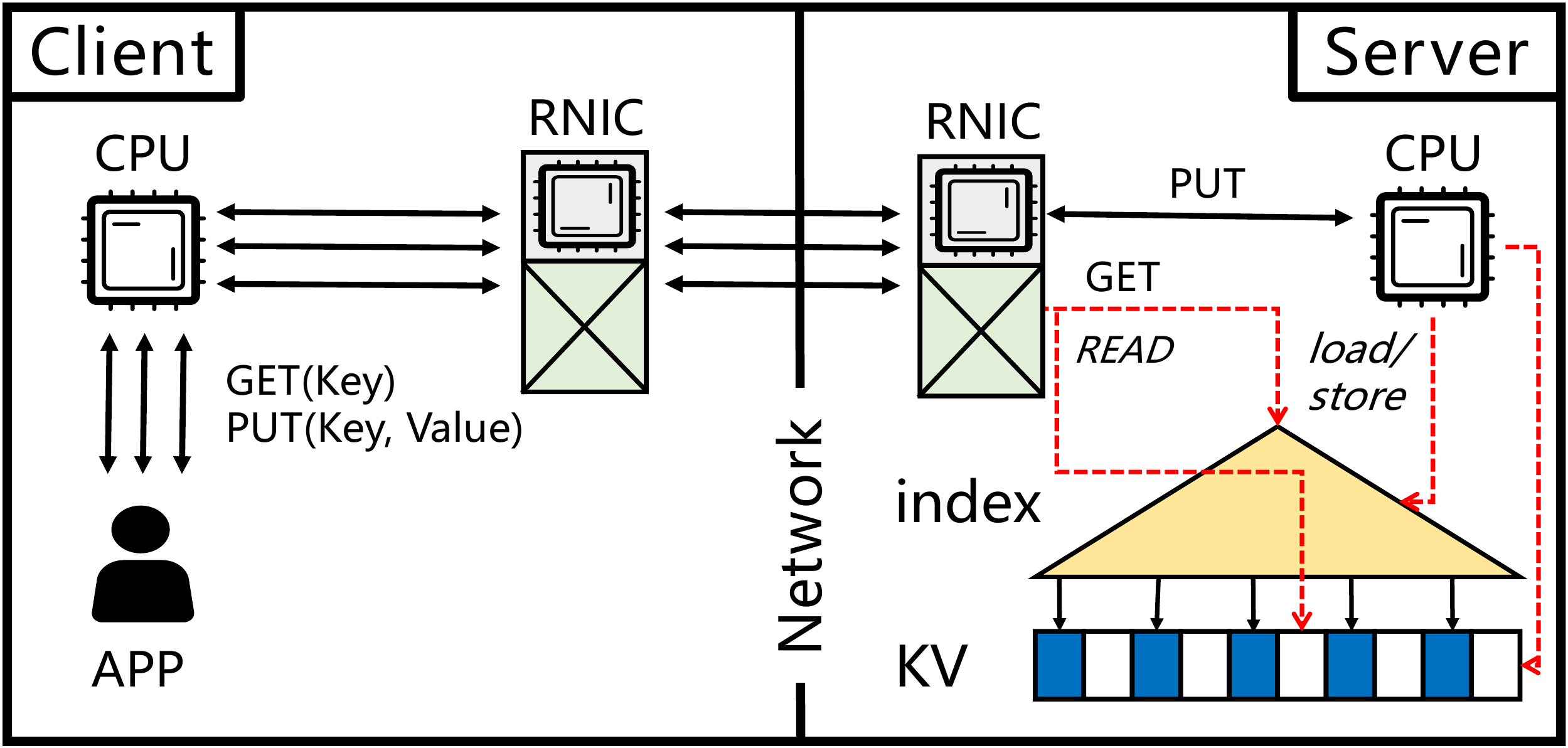}}
    \vspace{-4pt}
	\caption{\textbf{The architecture of different key-value stores with RDMA.}}
	\label{fig:rdma_kv}
\end{figure*}

\subsection{RDMA-based Key-Value Stores}

\noindent\textbf{RDMA Basics.}
RDMA is an alternative to network protocols (e.g., TCP, UDP), which allows fast data transfers
between local and remote memory with kernel bypassing~\cite{linux-rdma, kalia16rdma}.
RDMA hosts establish communication using \textit{queue pairs} (QPs) consisting of a send queue and a
receive queue, and post operations to the queues via different \textit{verbs}.
RDMA communications provide two types of verbs API, \textit{one-sided verbs} (aka \textit{memory
verbs}) and \textit{two-sided verbs} (aka \textit{message verbs}). The one-sided verbs, including
\texttt{READ}, \texttt{WRITE}, \texttt{CAS} (compare-and-swap), and \texttt{FAA} (fetch-and-add),
enable to directly access a pre-allocated memory region on a remote server without involving the remote
CPU. Two-sided verbs work like the conventional network protocols based on messages, where one
process sends/receives a message using the \texttt{SEND}/\texttt{RECV} verb.
Note that data transfer based on two-sided verbs incurs CPU cost on the remote server.

\noindent\textbf{Different architectures of RDMA-based key-value stores.}
A large number of research works~\cite{mitchell13pilaf, kalia14herd, wei15drtm, dragojevic15farm,
wang15hydradb, mitchell16cell, kalia2016fasst, cassell17nessie, kalia19erpc, ziegler19, wei20xstore,
zuo21race} have studied how to leverage RDMA to optimize key-value
stores in terms of their storage requirements and the characteristics of RDMA primitives. The
architecture of RDMA-based key-value stores can be classified into,
\textit{server-centric}, \textit{client-direct}, and \textit{hybrid-access} designs, based on the
usage of different RDMA verbs.
\begin{itemize}[leftmargin=*]
	\item Some works~\cite{kalia14herd, wei15drtm, kalia2016fasst, kalia19erpc, ziegler19} adopt server-centric design
		by replacing the communication layer (e.g., RPC) in a key-value store with RDMA primitives.
		Figure~\ref{fig:server-centric} shows the workflow of processing a request, where the
		client sends a request to the server via RDMA network and the server returns the response after
		processing the request locally. Such design depends only two RDMA operations, one for
		sending request and one for receiving response, which simplifies the implementation because it
		only requires adding an RDMA-enabled communication module. However, server-centric design
		still involves the remote CPU, which limits the scalability of the key-value stores and
		degrades the overall performance when the CPU becomes the bottleneck.
	\item To bypass the CPU of the remote server, some systems~\cite{cassell17nessie, zuo21race, wang22sherman} choose
		client-direct architecture, enabling the clients to directly access a pre-allocated memory
		region on the server. As shown in Figure~\ref{fig:client-direct}, the client directly
		fetches/writes the data from/to the server using one-sided verbs.
		Though the client-direct approach reduces the server CPU cost, it requires multiple
		network round trips to complete one complex operation, e.g., traversing a tree-based index on the
		remote server. Hence, the client-direct design incurs long latency when dealing with some
		complex data structures.
	\item Many existing works~\cite{mitchell13pilaf, dragojevic15farm, wang15hydradb, mitchell16cell,
		wei20xstore} leverage hybrid-access design to combine the advantages of server-centric and
		client-direct architectures. Figure~\ref{fig:hybrid-access} illustrates how the hybrid-access
		design works where the usage of one-sided verbs is restricted to read-only requests (i.e.,
		GET and SCAN). That is, the client can directly access the data from the remote server 
		for read-only operations, while the server only needs to process the requests
		involving writes (i.e., PUT, DELETE, and UPDATE).
		Thus, the hybrid-access designs not only harness the high performance of RDMA networks,
		but also relax the burden on the server CPUs.
\end{itemize}

\subsection{RDMA-friendly Index Structures}\label{sec:rdma_index}

\begin{figure*}[t]
	\centering
	\subfloat[Hash table]{
		\label{fig:hashtable}
		\includegraphics[width=0.3\textwidth]{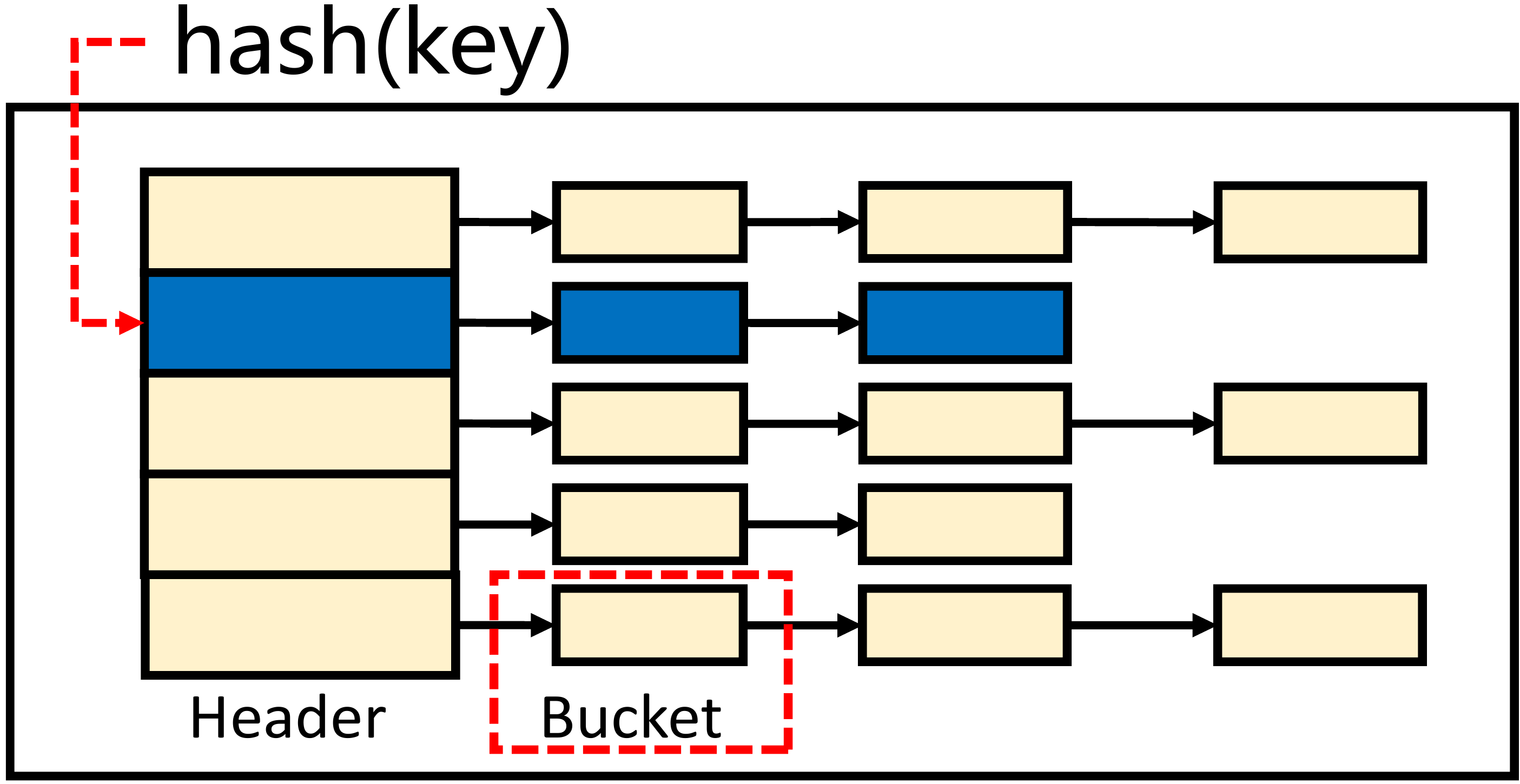}}
	\hspace{10pt}
	\subfloat[B$^{+}$-tree]{
		\label{fig:btree}
		\includegraphics[width=0.3\textwidth]{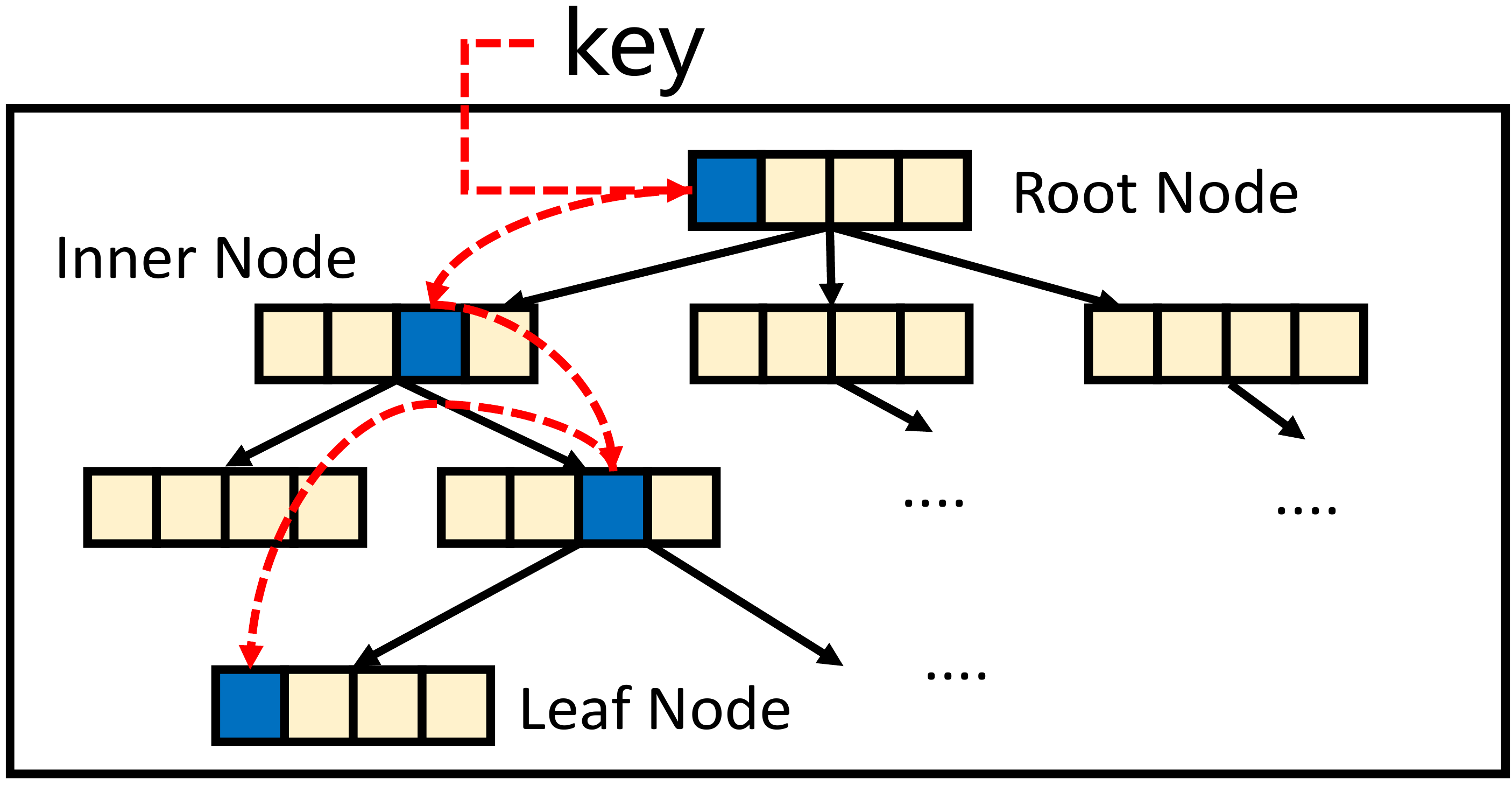}}
	\hspace{10pt}
	\subfloat[Skiplist]{
		\label{fig:skiplist}
		\includegraphics[width=0.3\textwidth]{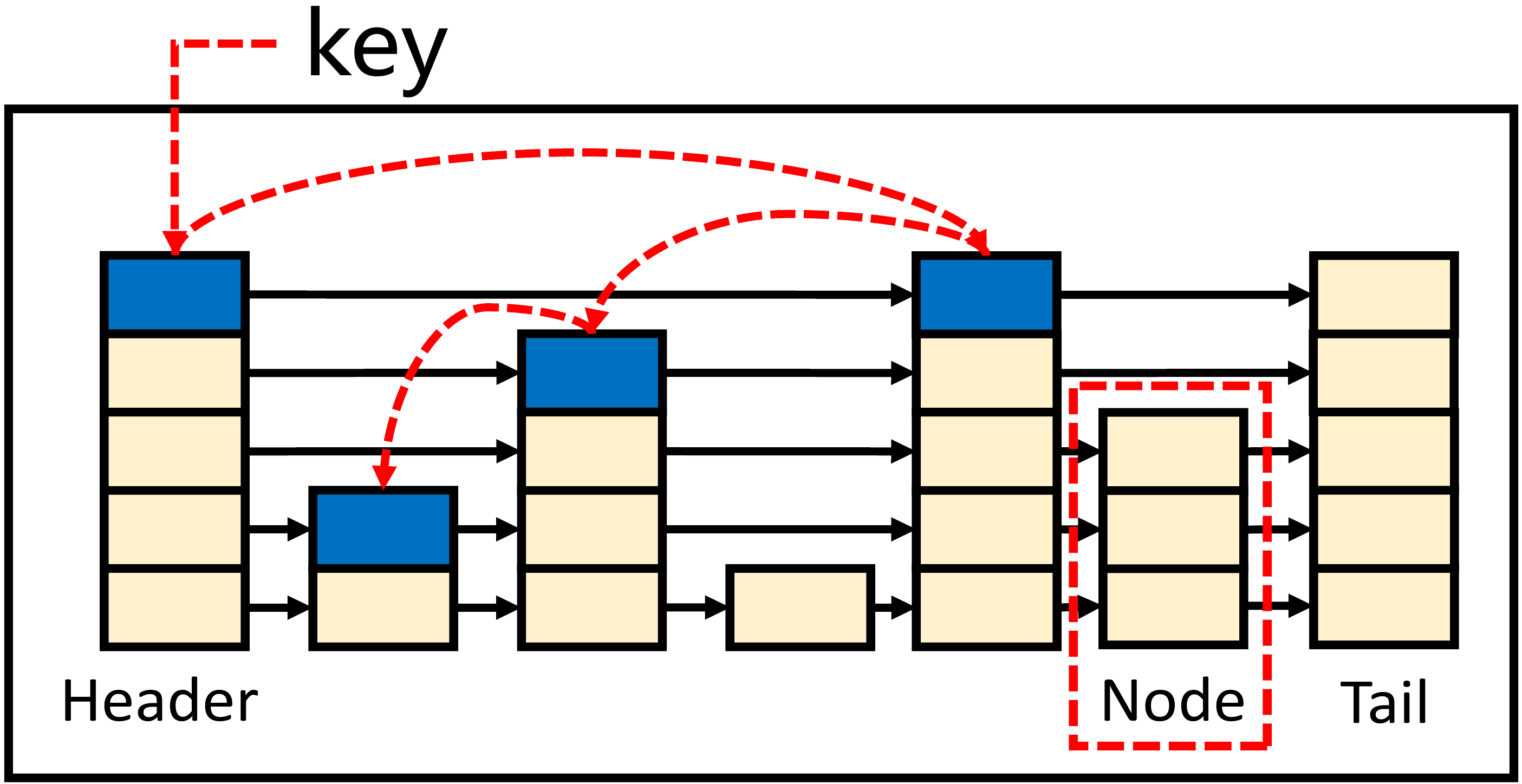}}
    \vspace{-4pt}
    \caption{\textbf{The data structure of different indexes.}}
	\label{fig:index}
\end{figure*}

\begin{figure*}[t]
	\centering
	\subfloat[Single-key lookup in DRAM]{
		\label{fig:memory_access}
		\includegraphics[width=0.24\textwidth]{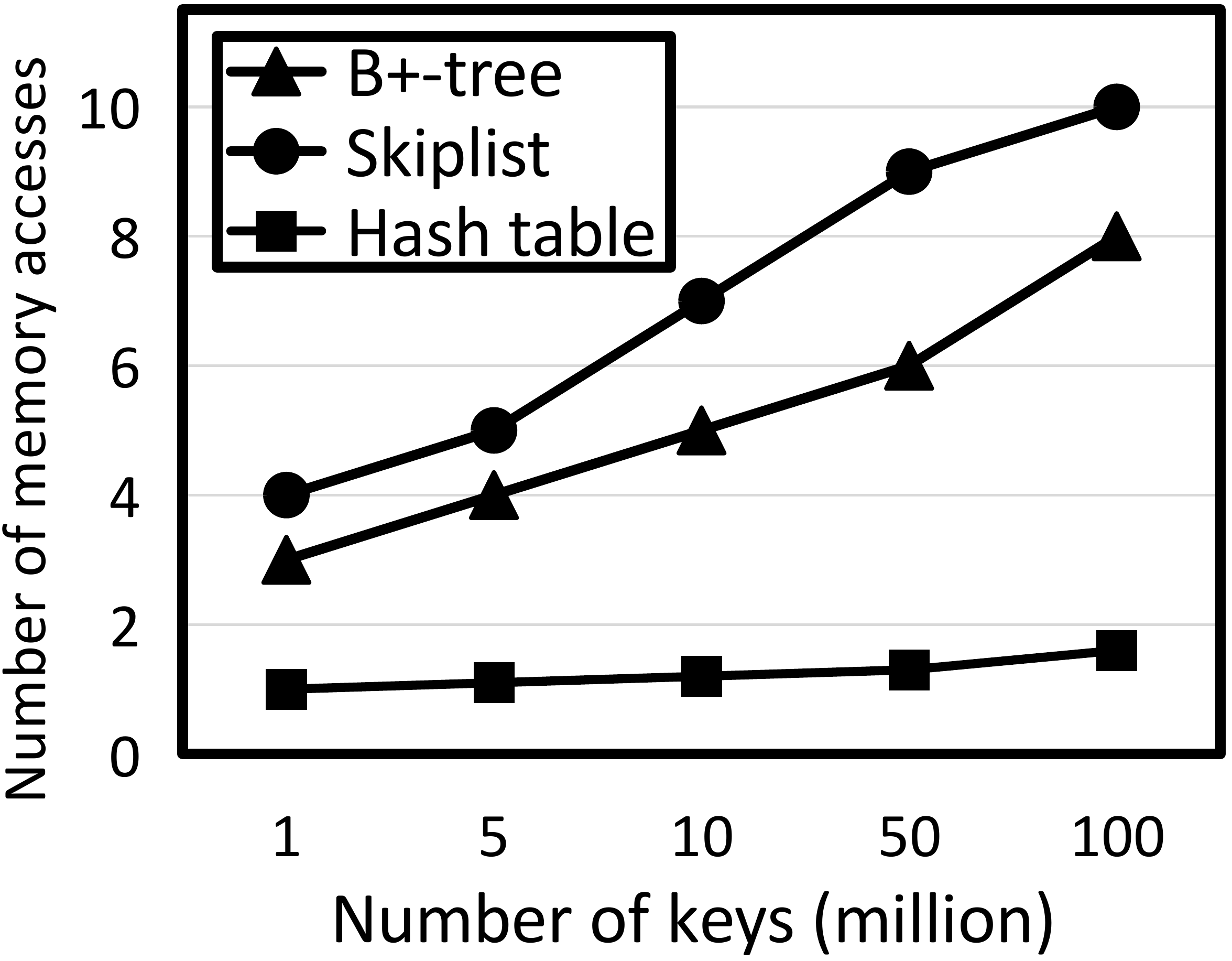}}
	\subfloat[Single-key lookup over RDMA]{
		\label{fig:index_latency}
		\includegraphics[width=0.24\textwidth]{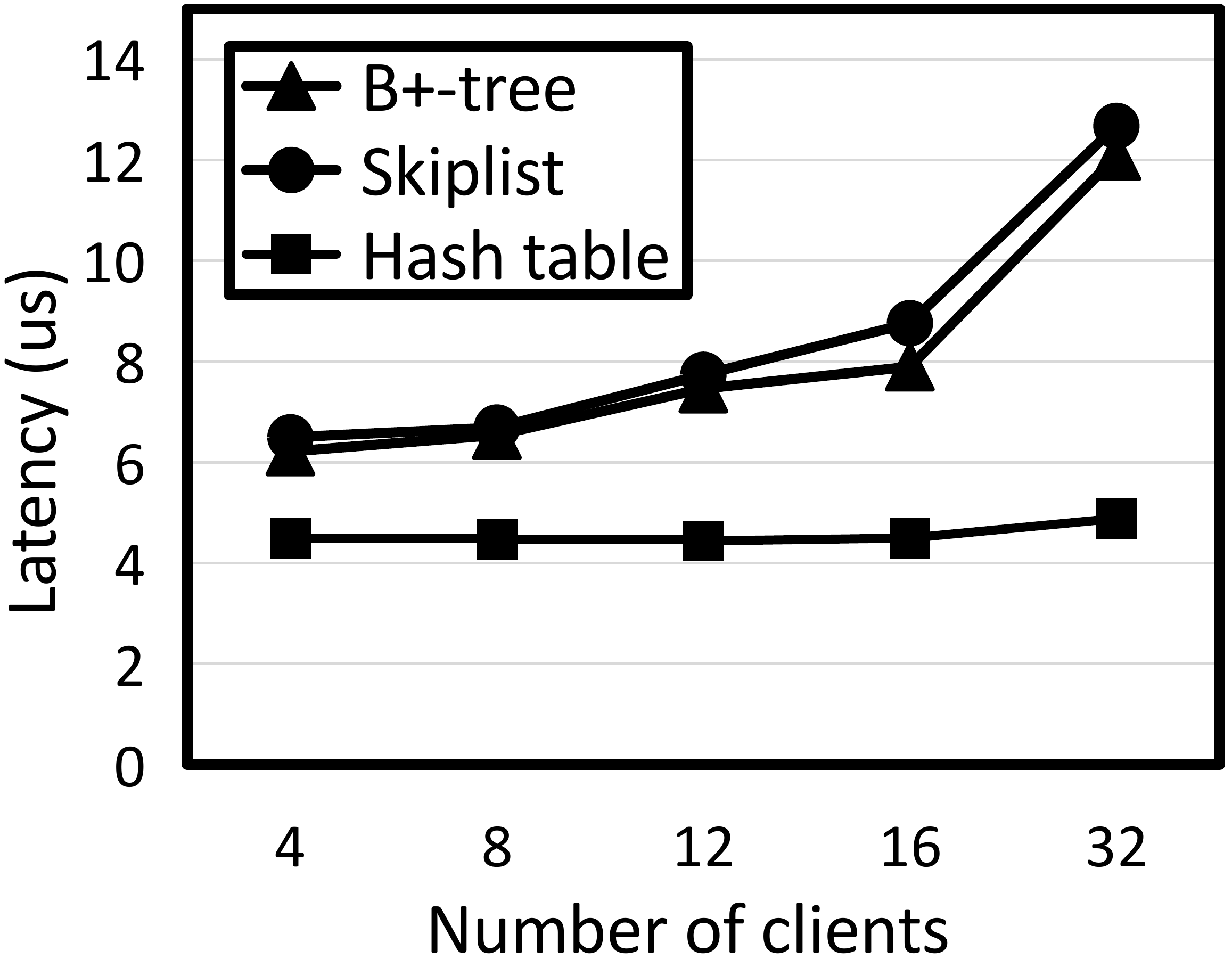}}
	\subfloat[PUT latency]{
		\label{fig:index_put}
		\includegraphics[width=0.24\textwidth]{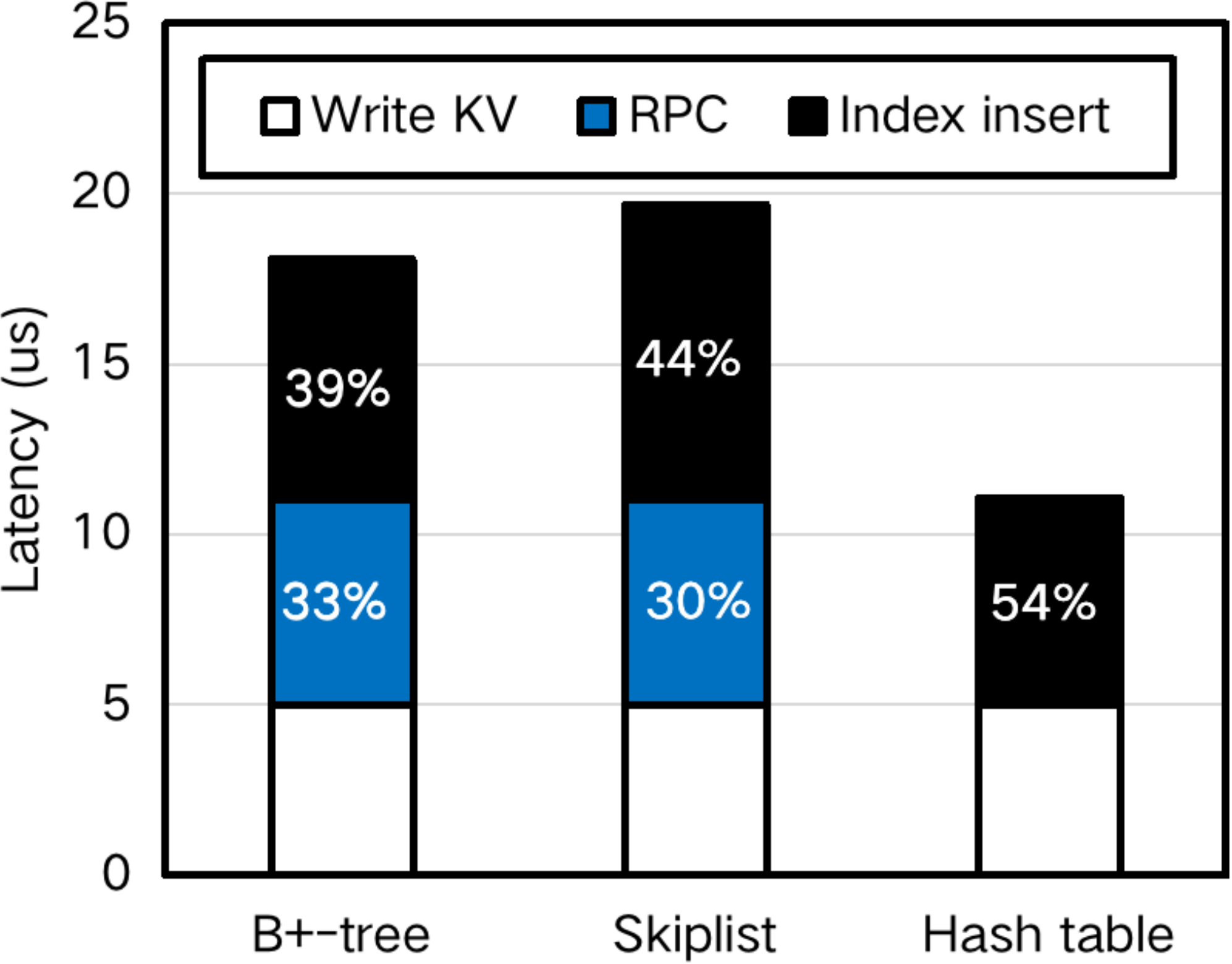}}
	\subfloat[GET latency]{
		\label{fig:index_get}
		\includegraphics[width=0.24\textwidth]{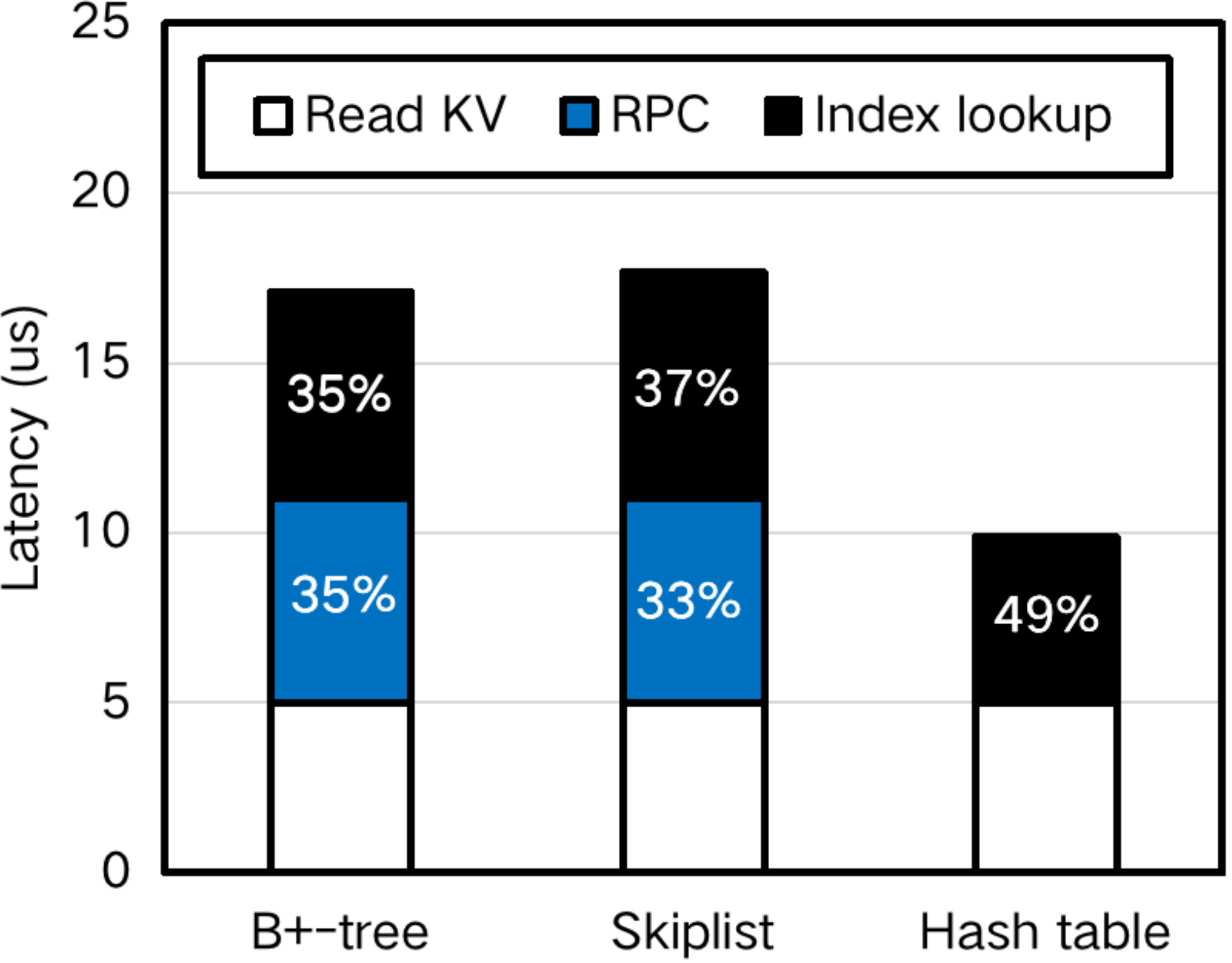}}
    \vspace{-4pt}
	\caption{\textbf{The performance of different index structures.} Note that the number of memory
	accesses in (a) equals to the number of round trips when we use one-sided verbs for the
	index without any optimization.}
	\label{fig:rdma_index}
\end{figure*}

As a key component of key-value stores, the indexes can determine the overall operation performance
(i.e., the overall time of a key-value operation).
Prior works have proposed many efficient index structures for key-value stores with
RDMA. Here we summarize the characteristics of different index structures, either hash-based or sorted (e.g.,
tree-backed, skiplist), and the applications of the indexes in RDMA-based key-value stores.

\noindent\textbf{Hashing index.}
Hashing index has been widely adopted in the RDMA-enabled key-values stores~\cite{mitchell13pilaf,
dragojevic14farm, kalia14herd, wang15hydradb, wei15drtm, kalia2016fasst, cassell17nessie, zuo21race},
because the hashing index can provide fast lookup services. Figure~\ref{fig:hashtable} shows the
data structure of a basic hash table, where a value is indexed by the hashing value of a key.
The simple data structure of the hash table enables the clients to directly fetch data
from the remote server using one-sided RDMA primitives (i.e., client-direct or hybrid-access design),
which mitigates the CPU overhead on the server. Thus, the key-value stores can achieve high
performance for single-point operations (e.g., GET, PUT, UPDATE, DELETE) using hashing index.
However, the hashing index does not support range queries, i.e., SCAN operations, limiting its
applications in building key-value stores.

\noindent\textbf{Sorted index.}
To support efficient range queries, many RDMA-based key-value stores~\cite{dragojevic15farm,
mitchell16cell, chen16drtmr, kalia19erpc, ziegler19, wei20xstore} leverage sorted index (e.g., B$^{+}
$-tree, skiplist), which organizes the key-value pairs in an ordered manner. Figure~\ref{fig:btree}
and \ref{fig:skiplist} illustrate the data structure of B$^{+}$-tree and skiplist respectively.
When locating a key-value pair, the sorted index requires multiple lookups.
For example, the B$^{+}$-tree needs to search from the root node to the leaf nodes, while the
skiplist requires multiple random accesses among its nodes. Note that the number of searching
operations increases with the data amount as the tree or skiplist grows larger.
Also, the complex structure of sorted index makes it complicated to update the index during writes. 
Hence, the RDMA-based key-value stores typically use two-sided verbs (i.e., server-centric or
hybrid-access design) to deal with sorted index, which reduces the number of communication round
trips but incurs server CPU cost. 

\subsection{Motivation and Challenges}
\label{subsec:motivation}

We conduct some experiments with RDMA communications to compare the indexing performance (e.g.,
index lookup, index insert) of
different index structures. We use two machines equipped with two Intel Xeon Gold 5215 CPU (2.5\,
GHZ), 64\,GB memory, and one 100\,Gbps Mellanox ConnectX-5 Infiniband NIC, to be a key-value store
server and a client respectively. The server and client are connected by a 100\,Gbps switch.
We set the key/value size to be 16\,B/32\,B respectively according to previous study~\cite{caofast20}.

\noindent\textbf{Observation~\#1: The RDMA-based systems using sorted index should leverage two-sided
verbs if possible, to minimize the number of round trips without any specific optimization.}
Although some recent works \cite{ziegler19, wang22sherman} explore to support sorted index
efficiently with pure one-sided verbs (to deploy on disaggregated memory), the optimized read and
write operations still need multiple round trips. For example, 94.1\% of write operations need at
least three round trips in Sherman \cite{wang22sherman}, a state-of-the-art distributed B$^{+}$-tree
optimized for writes. For read operations, both FG \cite{ziegler19} and Sherman require to cache
index locally on the client to avoid traversing the tree nodes, but there still remain some index
lookup operations with read retries (even experience nine times).

We compare the performance of different index structures by measuring the \textit{number of memory
accesses} on a key-value server, which equals to the amount of round trips when we adopt one-sided
verbs for the index without any optimization (e.g., combining dependent RDMA commands).
We test three common index structures (i.e., B$^{+}$-tree \cite{stxB+tree}, chained hash table, and
skiplist \cite{skiplist_github}) respectively, by accessing each key one by one using a single
lookup thread. All indexes reside in the memory.
Figure~\ref{fig:memory_access} shows the number of memory accesses of sorted index and hashing index
under different data amounts (1 to 100\,millions key-value pairs).
With larger number of key-value pairs, the number of memory accesses of B$^{+}$-tree and skiplist,
increases from 3 to 10, because B$^{+}$-tree expands itself by adding more nodes while skiplist
splits itself into more lists. For hash table, the number of memory accesses remains relatively
stable (around 1) with slight increase as more buckets are accessed in case of hash conflicts.
Our experiment results confirm the previous analysis that using one-sided verbs for sorted index
requires several round trips for one single-key search.
Therefore, if there is no constraint on the choices of RDMA commands (e.g., in case of disaggregated
memory), we should employ two-sided verbs to support sorted index efficiently without modifying the
data structures and adding other optimizations.

\noindent\textbf{Observation~\#2: The index lookup performance of sorted index drops down a lot with
more clients, as the server CPU becomes the bottleneck.}
We evaluate the {\it indexing latency} (time of index lookup over RDMA) of different indexes under different number of client
threads. Based on observation~\#1, we implement B$^{+}$-tree and skiplist using
eRPC~\cite{kalia19erpc} (an RPC library based on two-sided verbs), and access hash table via
one-sided verbs.
Here we load 100\,millions keys to each index structure, and start different number of client
threads issuing GET requests over all keys uniformly.
Figure~\ref{fig:index_latency} illustrates the indexing latency of different index structures under
various number of clients. Here the indexing latency includes the RDMA transfer time and the key
lookup time.  For the sorted index (i.e., B$^{+}$-tree, skiplist), the indexing latency increases
with the number of clients as the server CPU fails to handle all the client requests in time. For
the hashing index, the indexing latency remains relatively stable across different number of
clients because there is no CPU cost through one-sided verbs.
The indexing latency of sorted index is almost three times of that of hash table when there are 32
client threads issuing requests to the server.
Therefore, hashing index provides higher performance than sorted index, which motivates our idea of
combining hashing index and sorted index in RDMA-based key-value stores. 

\noindent\textbf{Observation~\#3: The indexing latency accounts for a large proportion of the
overall operation time.}
We further quantify the percentage of indexing latency in the {\it overall operation time} (the
total time of a complete PUT/GET operation), which includes index insert/lookup and value
insert/fetch over RDMA.
We start 32 client threads to write/read 100 millions key-value pairs uniformly.
Figure~\ref{fig:index_put} and~\ref{fig:index_get} depict the percentage of indexing latency in the
overall latency of PUT and GET operations respectively, where the latency of writing/reading KV
means the time to write/read a key-value pair using one-sided verbs once the index identifies the value
location. For the sorted index, the indexing latency includes the time of index insert/lookup and
network transfer through RPC. As we access the hash table using one-sided verbs, the indexing
latency equals to the time of index insert/lookup shown in the figure (without RPC).
The indexing latency of B$^{+}$-tree and skiplist accounts for 70-74\% of the whole process, while the
indexing latency with hash table is about half of the total time. Hence, reducing the indexing latency
with hybrid index can significantly improve the overall performance of a key-value store based on RDMA.

However, it is non-trivial to realize the idea of hybrid index efficiently in RDMA-based key-value stores.
The first challenge is to preserve strong consistency between the hashing index and the sorted index,
the index structures and the key-value items. Also, if the system replicates the index and data for
fault tolerance (e.g., using three-way replication), the replicas should be consistent with each
other. Thus, we should carefully perform the synchronization between different index structures, as
RDMA networks can aggravate this problem (e.g., when updating the hybrid index, only one index
is updated successfully while the other fails to be updated due to network interruption).
Second, though the read-only operations benefit from the hybrid index, the write performance
drops down because the system needs to maintain two index structures and update both of them for
write-operations. How to mitigate the overhead of keeping and updating hybrid index remains
an open problem.
Last but not least, when failure occurs (e.g., one index server crashes), it is challenging to
support all key-value services efficiently and achieve fast index recovery based on the remaining
index structures.

%% file: 3_design.tex
\section{Design of \sysname}
\label{sec:design}

In this section, we first introduce an intuitive approach of hybrid index which is hard to
efficiently maintain strong consistency between different index structures. We then present our hybrid
index scheme and build a RDMA-enabled key-value store called \sysname using hybrid index.

\input{3_design_1}

\subsection{Hybrid Index Scheme}

\subsubsection{Index Groups}

\begin{figure}[t]
	\centering
	\includegraphics[width=0.8\linewidth]{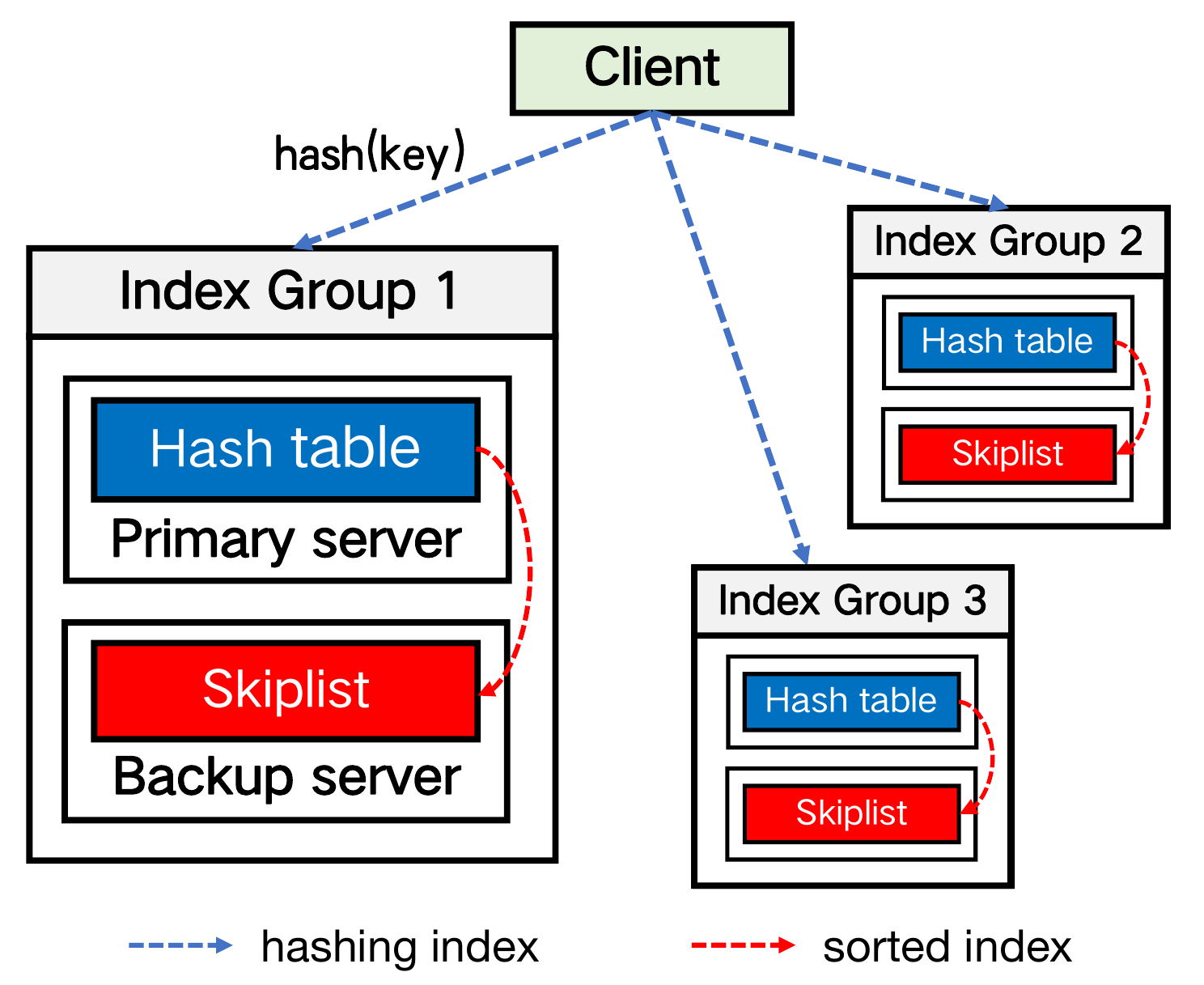}
    \vspace{-4pt}
	\caption{\textbf{Index groups.} \textit{We show three index groups here, and each index group
	consists of a hash table and a skiplist.}}
	\label{fig:index_groups}
\end{figure}

To efficiently manage the hybrid index with strong consistency, we propose \textit{index group} which
combines a part of hash table and a range of sorted index (e.g., a tree-backed index or a skiplist).
An index group is a unit of hybrid index.
The key idea here is to associate one index with another by index groups.
As shown in the intuitive approach, it is hard to update the two different index structures in an
efficient way, if there is no any relation between the hash-based index and the sorted index.
Hence, we map the sorted index to the hash table in the same index group, such that the indexes in
one group can get updated consistently.
This idea is similar to primary-backup replication in the hybrid transaction/analytical processing
(HTAP) systems~\cite{shen2021vegito}, where the OLTP and OLAP workloads run on primary and backup
replicas separately.
Following the naming of primary-backup replication, we call the index server managing hash table as
\textit{primary server} and the index server keeping skiplist as \textit{backup server}, because
GET operations which will be served by the hash table dominate in
real-world key-value workloads~\cite{atikoglu2012workload}.

Figure~\ref{fig:index_groups} depicts how the index groups combine two different indexes.
Here we show three index groups consisting of a hash table and a skiplist. As each index group is
responsible to manage a set of hash values, the index groups are independent with each other.
When inserting one key, it is first allocated to an index group based on its hash value, and then
inserted to the skiplist with two-sided verbs to keep the order of the keys.
For example, a key "brand" belonging to the second index group is inserted to the hash table of index
group 2 and the corresponding skiplist.

\subsubsection{Index Updating}

For write requests (PUT, DELETE, UPDATE operations), \sysname updates both the hash table and
the sorted index belonging to an index group to keep them consistent.
Compared to updating a hash table, updating a sorted index is a much more costly operation as it
involves additional operations to maintain the order of the keys. For example, inserting one key
into a B$^+$-tree index may trigger node splitting and merging while updating a skiplist may require
list splitting. Thus, \sysname performs asynchronous updates to the sorted index like
HiKV~\cite{xia17hikv} to keep the write latency low, while synchronously updates the hash table for
single-key lookups.

In case of index server failures, \sysname stores the updates in an append-only log before applying
the updates to the index structures.
Each log entry consists of a key, a value address, and a mark named "isApplied" to indicate whether
this key has been inserted to the local index structure.
When receiving a write request from a client, the primary server first records the update in its local log and
sends the update to the backup server; the backup server stores the update in its log, replies
to the primary server, and asynchronously applies the update to the skiplist; after receiving
successful response from the backup server, the primary server updates its hash table and returns
to the client.
As the sorted index is updated asynchronously, the backup server updates the index based on its log
before answering SCAN requests for strong consistency.
In other words, \sysname supports serializability that the written items can always be
accessed during single-point reads and range queries.

\subsubsection{Consistency Guarantee}

\sysname maintains strong consistency from two perspectives: (i) consistent index structures in each
index group, and (ii) consistent index with the key-value data.
For the first consistency issue, our approach to update the index based on the log can keep the sorted
index consistent with the hash table in the same group. A complete index update means that the
update has been recorded in the logs on both the primary server and the backup server, and the hash
table gets updated. As the log on the primary server and the backup server is the same, the sorted
index is consistent with the hash table when the sorted index finishes applying the updates
asynchronously.
For the second consistency point, it means that the data can be indexed during reads after the data
is successfully written to the key-value store. That is, if a write fails, the invalid key-value
pair should not be indexed during the reads. 
To ensure the index structures consistent with the data stored, \sysname uses sequential
writes to store the value and update the index.
When performing a write request, the client first stores the key-value pair to a data server and
gets the value address from the data server; then chooses an index group based on the hash value of
the key, and connects to the primary server for index update. If any step fails during the write
process, the write operation fails and the client can restart a write operation.

\subsubsection{Choices of RDMA Verbs}

\sysname uses different RDMA verbs for different index structures to achieve low latency.
For single-key lookups, \sysname allows the client to directly access the hash table on the primary
server using one-sided verbs, which bypasses the primary server CPU.
For range queries, the client sends the request to the backup servers maintaining sorted index via
two-sided verbs, which can be completed in one round trip. Upon receiving SCAN requests, the backup
server searches its local skiplist and returns the results to the client.
\sysname leverages two-sided verbs to deal with write requests, because the client requires to get
response from the primary server and the primary server needs to receive the response from the
backup server. 
In summary, \sysname uses one-sided verbs for GET operations and two-sided verbs for other
operations.

\subsection{Fault Tolerance}
\label{sec:tolerance}

\begin{figure}[t]
	\centering
	\subfloat[Double hash tables]{
		\label{fig:backup1}
		\includegraphics[width=0.22\textwidth]{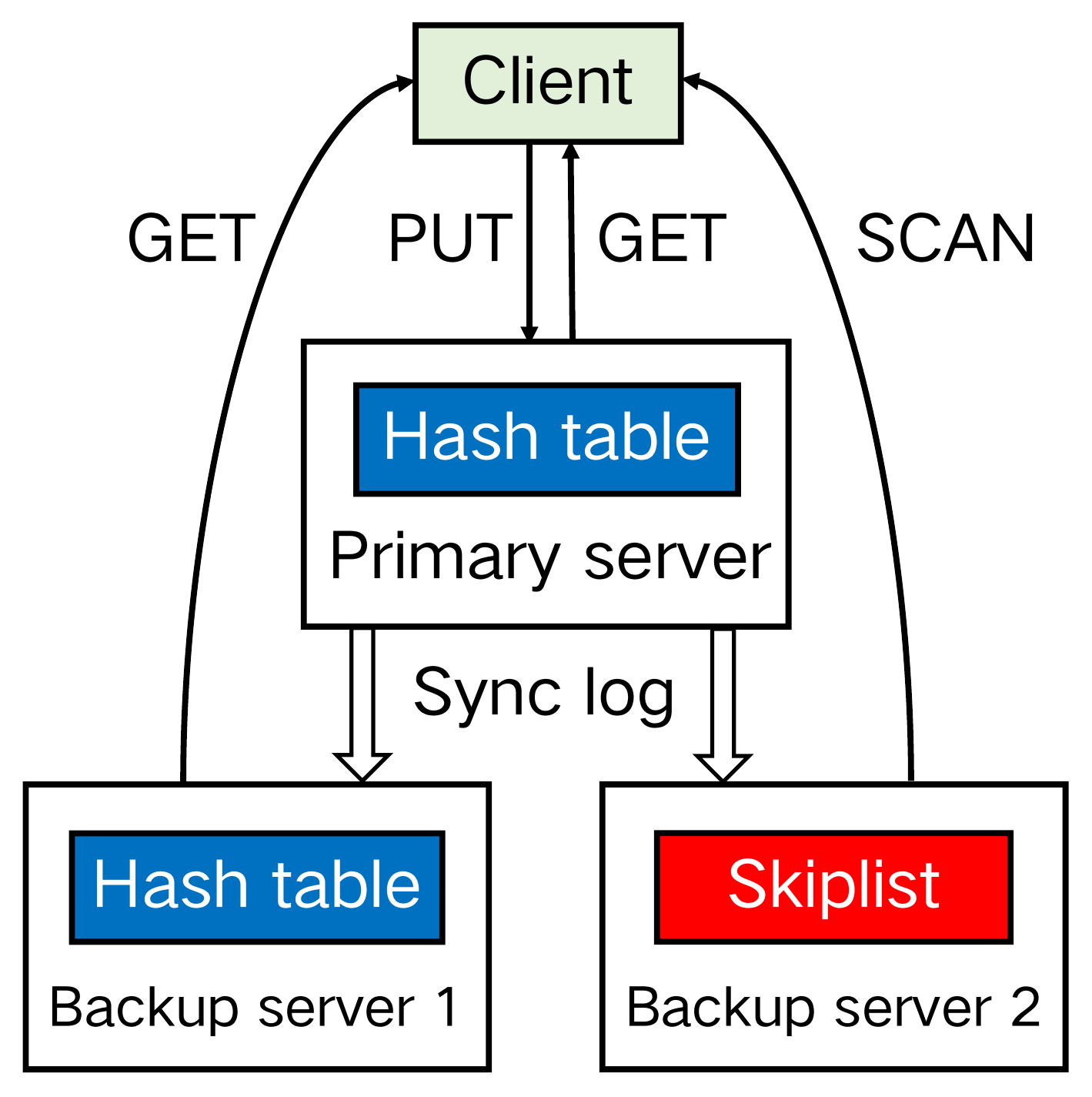}}
	\subfloat[Double skiplists]{
		\label{fig:backup2}
		\includegraphics[width=0.22\textwidth]{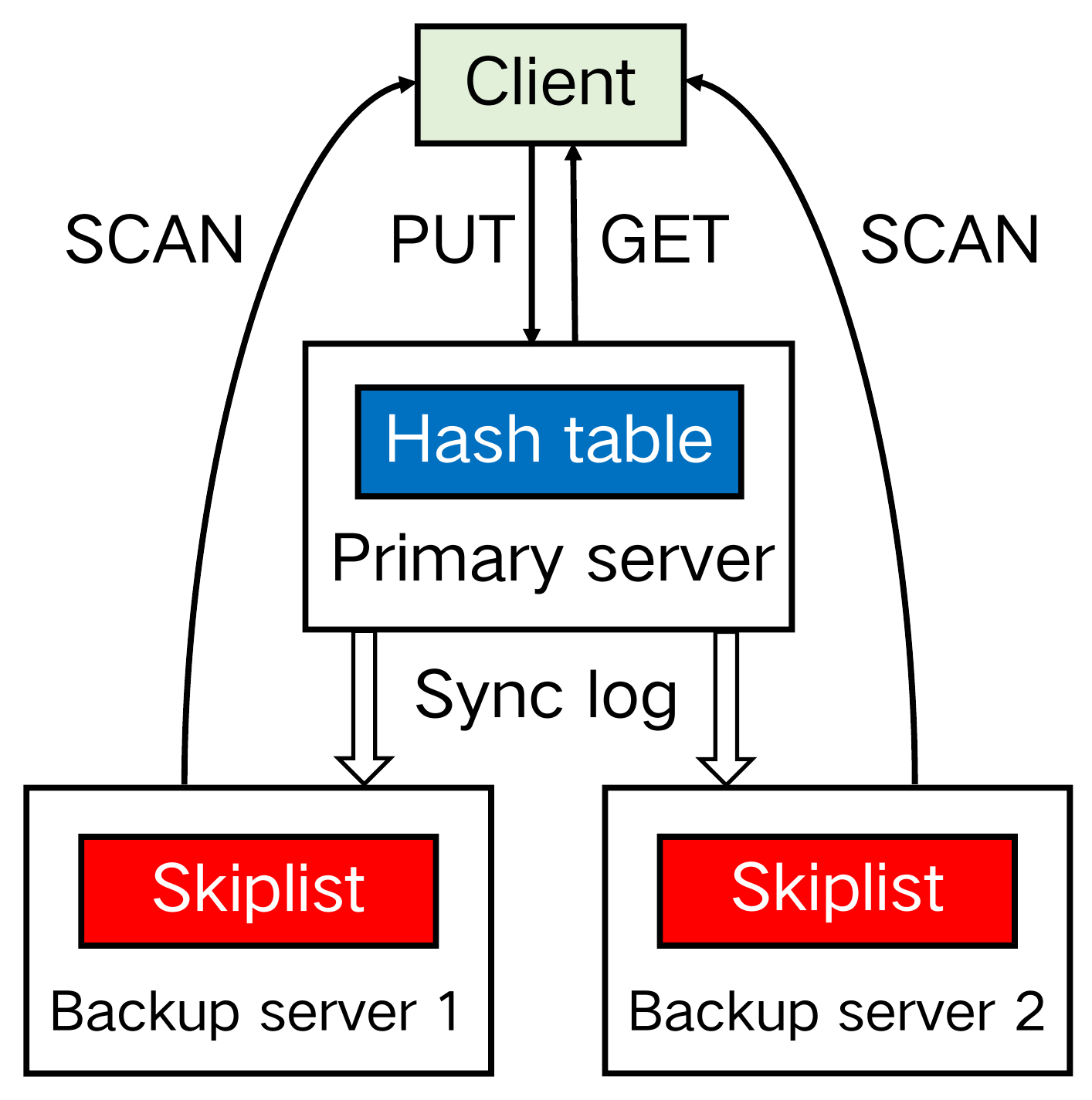}}
    \vspace{-4pt}
    \caption{\textbf{Two backup strategies for hybrid index when using
    three-way replication.} \textit{\sysname adopts the second strategy to
    achieve high availability.}}
	\label{fig:backup}
\end{figure}

Our basic approach of hybrid index cannot tolerate index server failures as an index group consists
of only one hash table and one sorted index. Though the sorted index can be considered as a backup
of the hash table, neither the hash table nor the sorted index has an exact copy. That is, only a
hash table or a skiplist remains when one index server in a group fails. If the
primary server fails, the backup server keeping the skiplist requires to provide single-key lookups,
which increases the latency of GET operations. In case of the backup server failure, the remained hash
table cannot support SCAN operations until the completion of rebuilding the sorted index. Thus, we should
address the issue of fault tolerance in hybrid index for high availability.

We consider adding one backup index structure to a group which basically consists of a hash
table and a sorted index, as distributed systems typically use three-way replication to achieve high
availability in the presence of failures~\cite{ghemawat2003gfs, digeo17raft}.
The added index can be a hash table or a sorted index. Figure~\ref{fig:backup} shows the two
possible backup strategies for hybrid index.
If we replicate the hash table in one additional backup server, there are two same hash tables and
one skiplist in an index group. As shown in Figure~\ref{fig:backup1}, the primary server and one
backup server manage two hash tables, and the other backup server keeps the skiplist. 
Thus, when the primary server fails, the backup server with the hash table can directly act as the
primary server. Moreover, the backup server containing the hash table can also provide single-point
lookups, which relieves the request burden (for GET operations) on the primary server in some
degree. However, this approach cannot tolerate the failure of the index server with skiplist.  
Another alternative is to replicate the skiplist on one additional backup server, which tolerates
the failure of one index server with skiplist but fails to tolerate the primary server failure.
When the primary server fails, one backup server with skiplist needs to support single-point
operations until the primary server is repaired.
Figure~\ref{fig:backup2} depicts such backup strategy where the primary server maintains a hash
table while two backup servers keep two same skiplists. In this case, both the two backup servers
with skiplist can support range queries.

\sysname replicates the skiplist in an index group by default, i.e., the primary server keeps a
hash table while two backup servers maintain skiplists. First, the hybrid index can provide rich
key-value operations in case of any index server failures in a group.
The failure of one backup server does not affect the functioning of the whole system, i.e., the hash
table on the primary server and the remained skiplist are sufficient to answer all requests.
If the primary server fails, the backup server can process single-point requests though the
GET performance reduces without the hash table.
Second, \sysname can quickly rebuild a hash table to recover the primary server, restarting to
support GET requests efficiently. As the recovery time of a hash table is much shorter than
that of a skiplist, we prefer to rebuild a hash table to reduce the impact of server failures on the
whole key-value store.
Third, the SCAN performance can benefit from one additional server because the management of
skiplist relies on the server CPU. If replicating hash table on one additional server, the
performance improvement of GET requests is limited, as the operations based on one-sided verbs do
not involve the server CPU.
Therefore, \sysname configures one hash table and two skiplists in one index group to achieve high
availability and low latency.

\begin{figure}[t]
	\centering
	\includegraphics[width=0.85\linewidth]{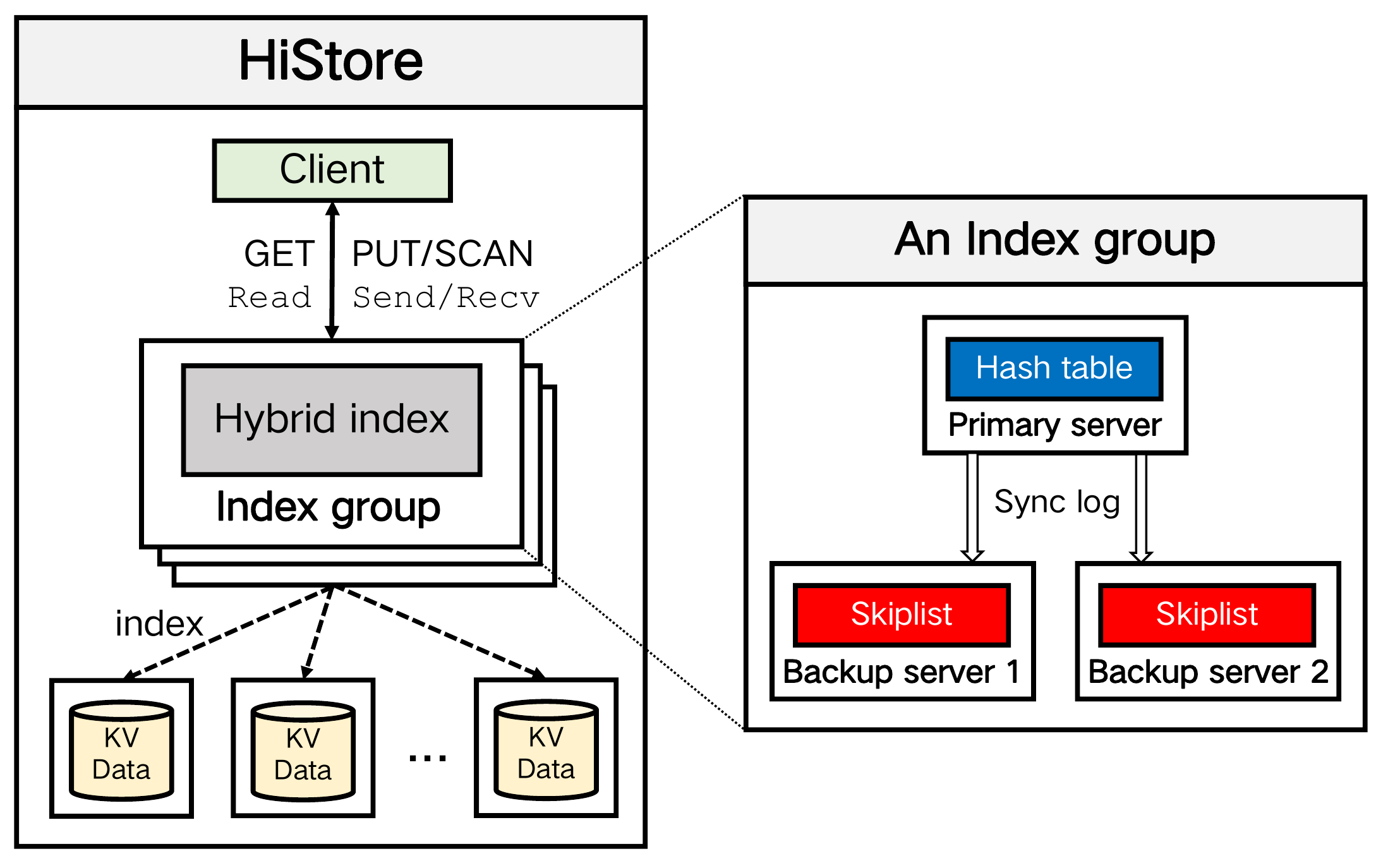}
    \vspace{-4pt}
	\caption{\textbf{The architecture of \sysname.} \textit{}}
	\label{fig:architecture}
\end{figure}

\noindent\textbf{Putting it all together.}
Figure~\ref{fig:architecture} shows the architecture of \sysname, which consists of multiple index
groups and many data servers. The index groups manage the hybrid index while the data servers are
responsible for storing the values.
In each index group, a primary server maintains a hash table for single-key lookups based on
one-sided verbs while two backup servers keep the skiplists for range queries using two-sided verbs.
During writes, the client first stores the values on the data servers, and then asks the primary server to update
the index using two-sided verbs.
\sysname asynchronously applies the update to the skiplists on the backup servers to improve the
write performance.

%% file: 3_design_1.tex
\subsection{An Intuitive Approach}
\begin{figure}[t]
	\centering
	\includegraphics[width=0.8\linewidth]{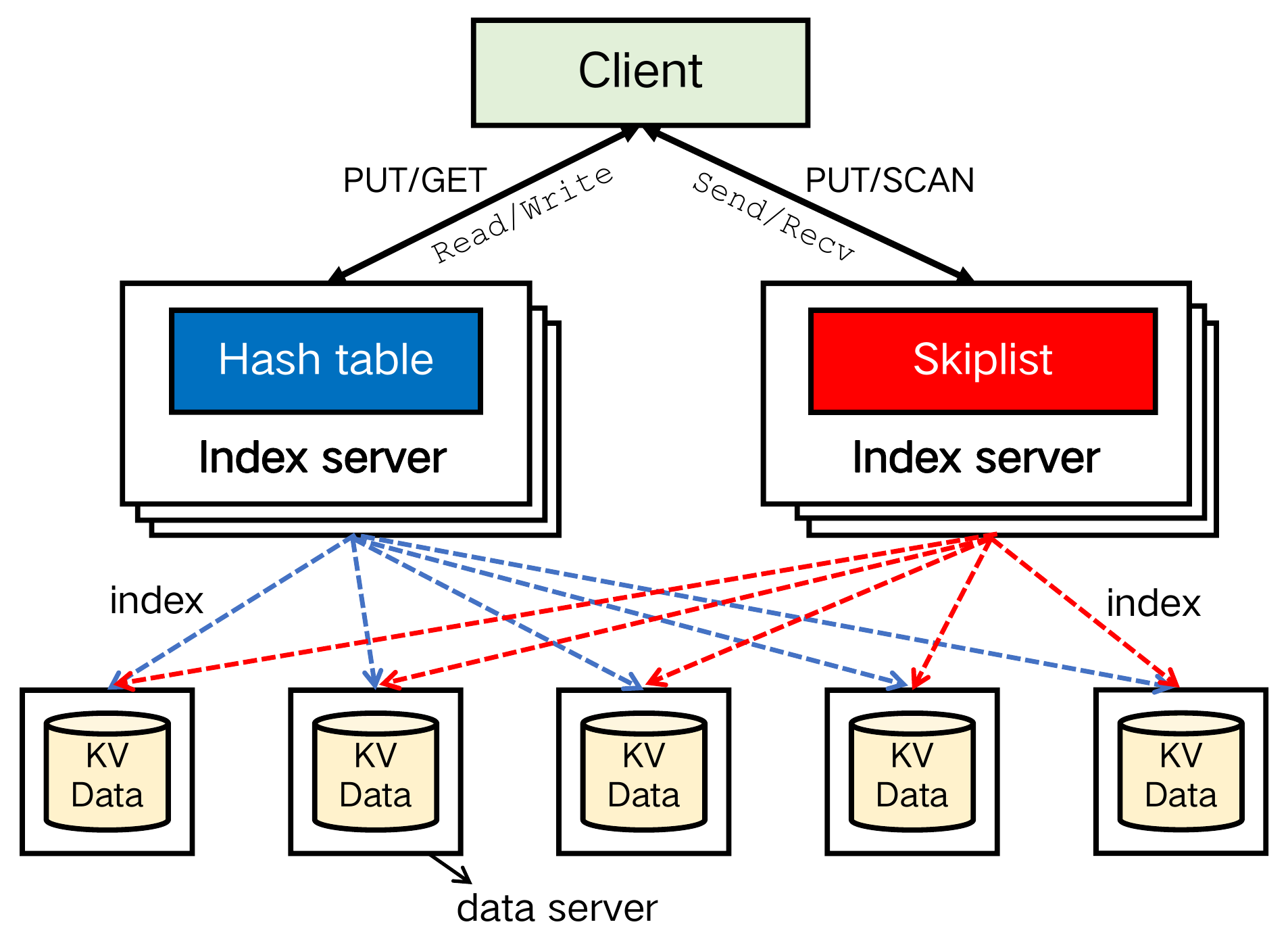}
    \vspace{-4pt}
	\caption{\textbf{An intuitive approach to build key-value store using hybrid index.} \textit{}}
	\label{fig:naive_approach}
\end{figure}

One natural approach to leverage hybrid index in RDMA-based key-value stores is using one-sided
verbs to access the hashing index while employing two-sided verbs to access the sorted index, driven
by our previous observations (in Section~\ref{subsec:motivation}).
Figure~\ref{fig:naive_approach} shows the intuitive design of building a key-value store
based on hybrid index, e.g., a hash table and a skiplist. To distinguish the server maintaining
index from that storing data, we call the server managing index as \textit{index server},
and the server storing values as \textit{data server}.
There can be multiple index servers and many data servers in a key-value store. Here we store the
hash table in some index servers and place the skiplist in the other index servers. 
The client chooses to access different index structures based on the type of read operations. That is, the
client directly accesses the hash table for GET operations while leveraging the
skiplist for range queries.
During writes, the client needs to update both the hash table and the skiplist using one-sided
and two-sided verbs respectively. 

Such approach seems to maximize the system performance, but poses unique challenges to maintain
consistency between the two indexes. One one hand, this design bypasses the server CPU as much as
possible (i.e., read/write the hash table using one-sided verbs) and minimizes the number of round
trips (i.e., access the skiplist via two-sided verbs). 
One the other hand, as the two indexes are totally separate, partitioning the data according to the
hash values of the keys helps address the issue of skewed access while the ranges of the key-value
pairs can be maintained by the sorted index.
However, it is hard to implement a lightweight mechanism to keep the hashing index consistent with
the sorted index in such a design. 
The root problem is how to atomically and efficiently update the two different indexes. 
It can be prevalent in distributed environment that one index is updated
successfully but the other one is not, caused by the network interruptions or server crashes.
Thus, the value can be indexed by one of the indexes or none of them.
One solution to the inconsistency problem is to employ distributed transactions~\cite{burrows06osdi}, which
requires to access the transaction table during writes. The system can judge the validity of an
update based on the transaction table, and then complete/roll back the whole operation. This
complicated method introduces additional overhead to the write process, degrading the write
performance.

%% file: 4_implement.tex
\section{Implementation}
\label{sec:impl}

We implement \sysname in C++, which realizes two-sided communications based on eRPC~\cite{kalia19erpc}.
In this section, we first introduce the threading model and data structures of hybrid index in
\sysname. We then explain how \sysname provides key-value services and recovers hybrid index
when failure occurs. 

\subsection{Hybrid Index}

\begin{figure*}[t]
	\centering
	\includegraphics[width=0.85\linewidth]{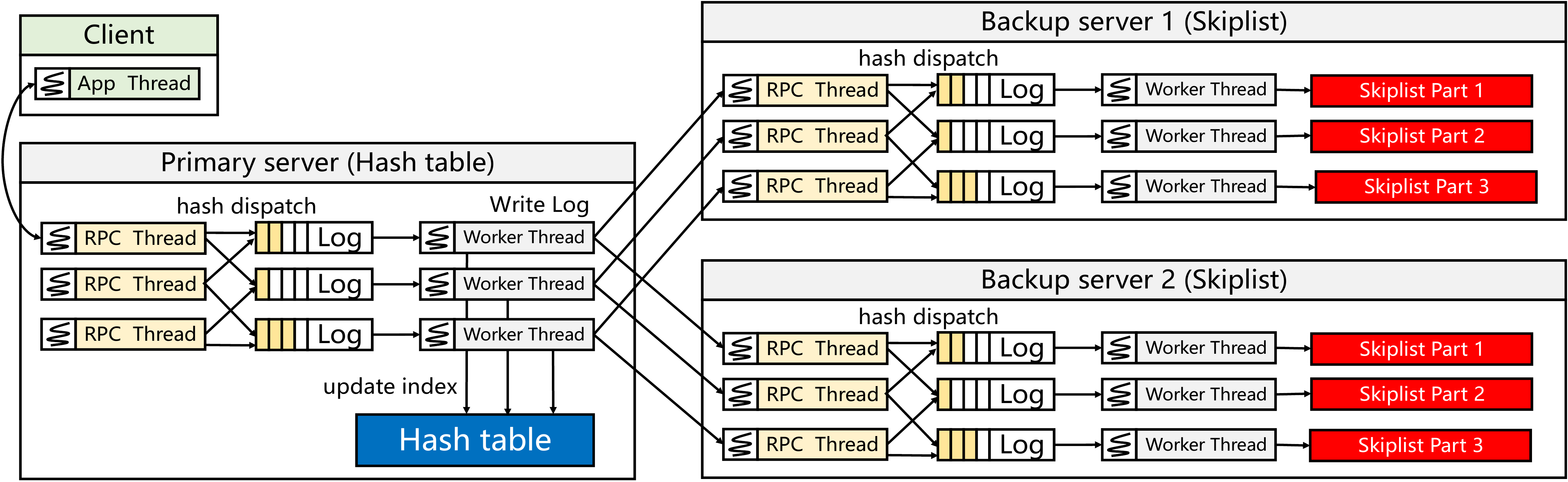}
    \vspace{-4pt}
	\caption{\textbf{The threading model of hybrid index in \sysname.}}
	\label{fig:thread}
\end{figure*}

\noindent\textbf{Threading Model.}
\sysname employs multi-threading model for writes and range queries which involve the server CPU.
Each index server starts several \textit{RPC threads}, which are responsible to receive and process
RPC requests, and \textit{worker threads}, which perform index updates and key lookups. 
Figure~\ref{fig:thread} depicts the threading model of hybrid index.
During writes, the worker threads on the primary server update the hash table while the worker threads
on the backup servers apply updates to the skiplist.
For range queries, the worker threads on the backup servers look up the keys among the sorted lists.
We explain how the RPC threads and worker threads support each key-value operation in
Section~\ref{subsec:key-value_services}.


\noindent\textbf{Data Structures.}
\sysname currently adopts a chained hash table and a basic skiplist to constitute an index group.
Note that the hash table and the skiplist can be replaced with other optimized hash table and sorted
index (e.g., tree-backed) respectively.
Each chain consists of multiple buckets, each of which is of 64\,B.
A bucket contains seven hash slots, and a pointer of 8\,B to link the next bucket.
Each hash slot records the information on a key-value pair, consisting of the hash value of the key
(1\,B), the length of the key-value item (1\,B), and the value address (6\,B).
For single-key lookups, the client first computes the bucket address based on the key locally,
reads a bucket using the \texttt{READ} verb, and then
searches the key within the bucket. The client needs to check each slot by comparing the signature
(i.e., the hash value of the key) and the exact key.
If the comparison succeeds, the value is returned. If no valid slot is matched and next pointer is not
empty, the next bucket is queried based on the next pointer.
When a bucket is full during writes (e.g., hash collision occurs), the client issues an RPC request
for adding a bucket to the server, which links a new bucket after the last bucket using the next
pointer; after receiving the successful response from the server, the client can rewrite the
key-value pair via one-sided verbs.
To avoid resizing (which introduces one more RDMA round trip), we allocate more buckets than
required by consuming a little more memory space.
For the skiplist, \sysname divides the whole list to several partitions based on the hash values of
the keys, such that each partition can be searched concurrently by multiple threads to reduce the
latency of range queries.

\subsection{Key-value Services}
\label{subsec:key-value_services}

\noindent\textbf{Write operations.}
\sysname uses two-sided verbs to perform index updating and data storing.
For PUT and UPDATE operations, the client first stores values on the data servers to get the
value addresses before updating the hybrid index.
Then the client sends requests for updating the index to the RPC threads on the primary
server.
The RPC threads on the primary server append the updates to different logs based on the hash values of
the keys. The worker threads then send the updates to the backup servers using two-sided verbs, and
wait for the responses from the backup servers.
To improve the write throughput, the worker threads perform log synchronization between the primary
server and the backup servers in a batch.
On the backup server, the RPC threads append the
updates to the log and send successful responses to the primary server, while the worker threads
asynchronously update the skiplist. As the skiplist is divided into several partitions, different
partitions can be updated by multiple worker threads concurrently.
Upon receiving the successful responses from the backup servers, the worker threads on the primary
server apply the updates to the hash table and return success to the client.
To handle concurrent writes and reads, the primary server updates the hash table by compare
and swap operation with CPU, which makes each update an atomic operation. If any step fails during
writes, the incomplete write operation is considered as a write failure; the client can restart the
write process if it does not receive successful response from the primary server after a period of
time.

\noindent\textbf{GET.}
\sysname handles GET requests using one-sided verbs, totally bypassing the server CPU. 
It leverages the hash table on the primary server for single-key lookups.
To read a key-value pair, the client directly accesses the hash table using one-sided verbs to
obtain the value address.
Then, the client retrieves the value from the data server according to the value address.
The whole process avoids incurring CPU cost on both the index server and the data server.

\noindent\textbf{SCAN.}
\sysname provides efficient range queries based on the skiplists.
The client submits SCAN requests via eRPC to the backup servers, where the worker threads search
among the skiplist partitions concurrently to return the results to the client. Note that the worker
threads make sure that none index updates remains before processing the query. That is, if there are
index updates left, the worker threads will first apply the updates to the skiplist and then answer
the SCAN request.

\subsection{Failure Handling}

When an index server fails, \sysname still provides key-value services and starts a recovery process
to rebuild the failed index on the new index server.
In case of the primary server failure, one backup server acts as the primary server temporarily to
process all writes while the other backup server handles the range queries; the GET requests is
distributed to any backup server randomly. To rebuild a hash table, the new primary server retrieves
the skiplist from the temporary primary server via eRPC. When the hash table is rebuilt completely,
the new primary server can handle key-value operations and the index group can work normally.
When a backup server fails, another backup server processes all SCAN requests. The new backup server
fetches the hash table from the primary server via eRPC to generate a skiplist, and works as a
normal backup server when its skiplist is up-to-date. 

%% file: 5_evaluation.tex
\section{Evaluation}
\label{sec:eval}

\begin{figure*}[ht]
	\centering
	\subfloat[PUT operation]{
		\label{fig:basic_put_tp}
		\includegraphics[width=0.30\textwidth]{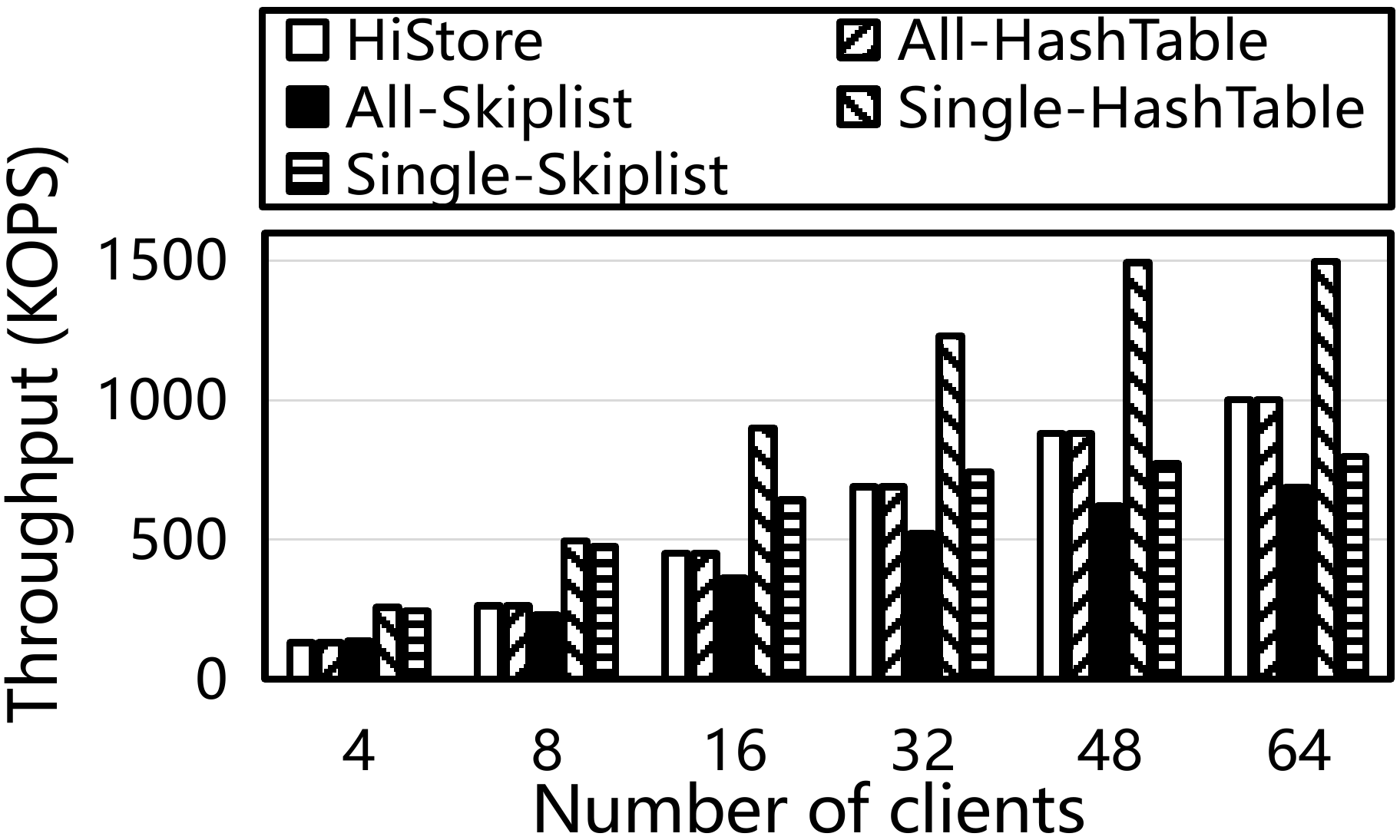}}
	\hspace{10pt}
	\subfloat[GET operation]{
		\label{fig:basic_get_tp}
		\includegraphics[width=0.30\textwidth]{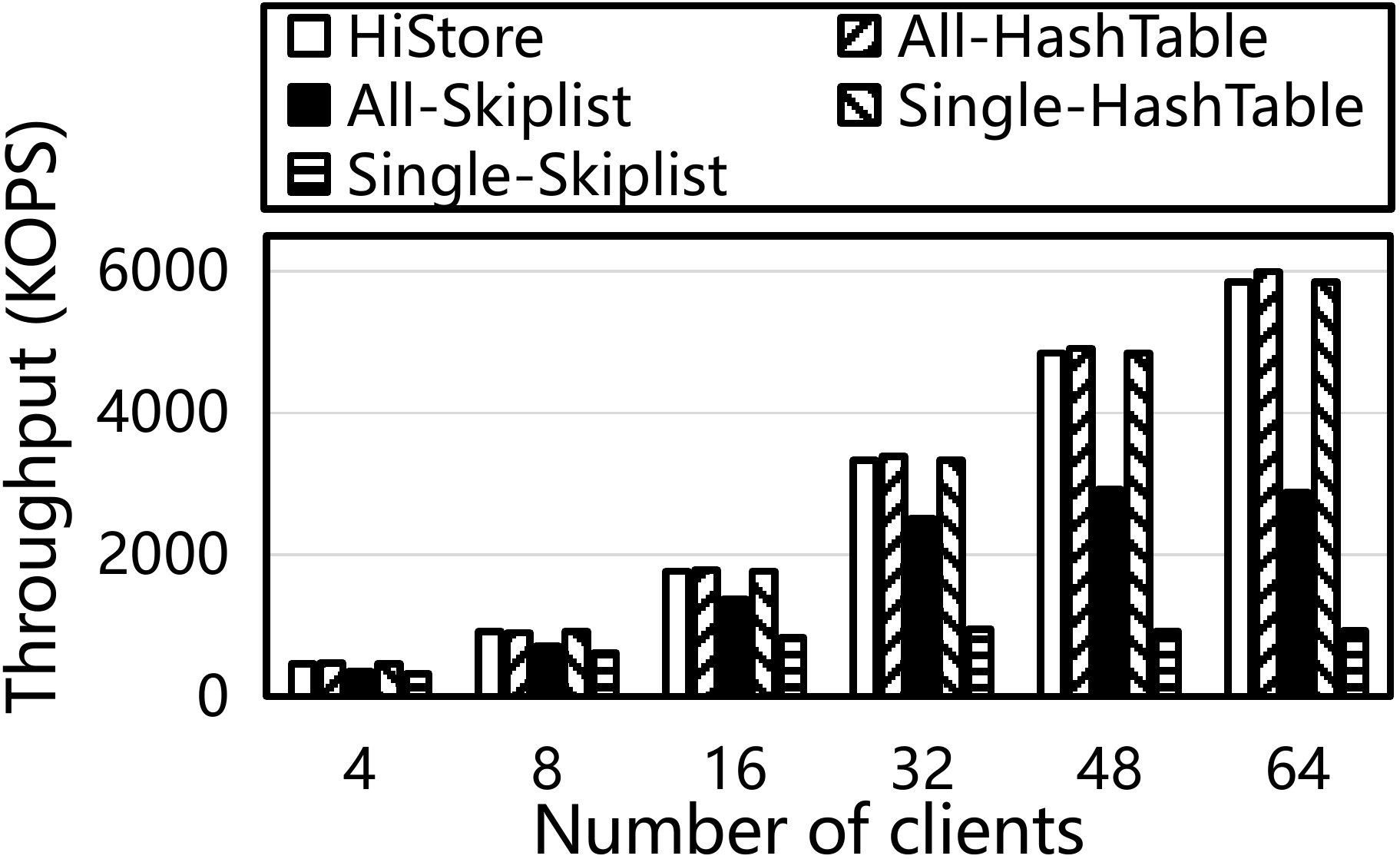}}
	\hspace{10pt}
	\subfloat[SCAN operation]{
		\label{fig:basic_scan_tp}
		\includegraphics[width=0.30\textwidth]{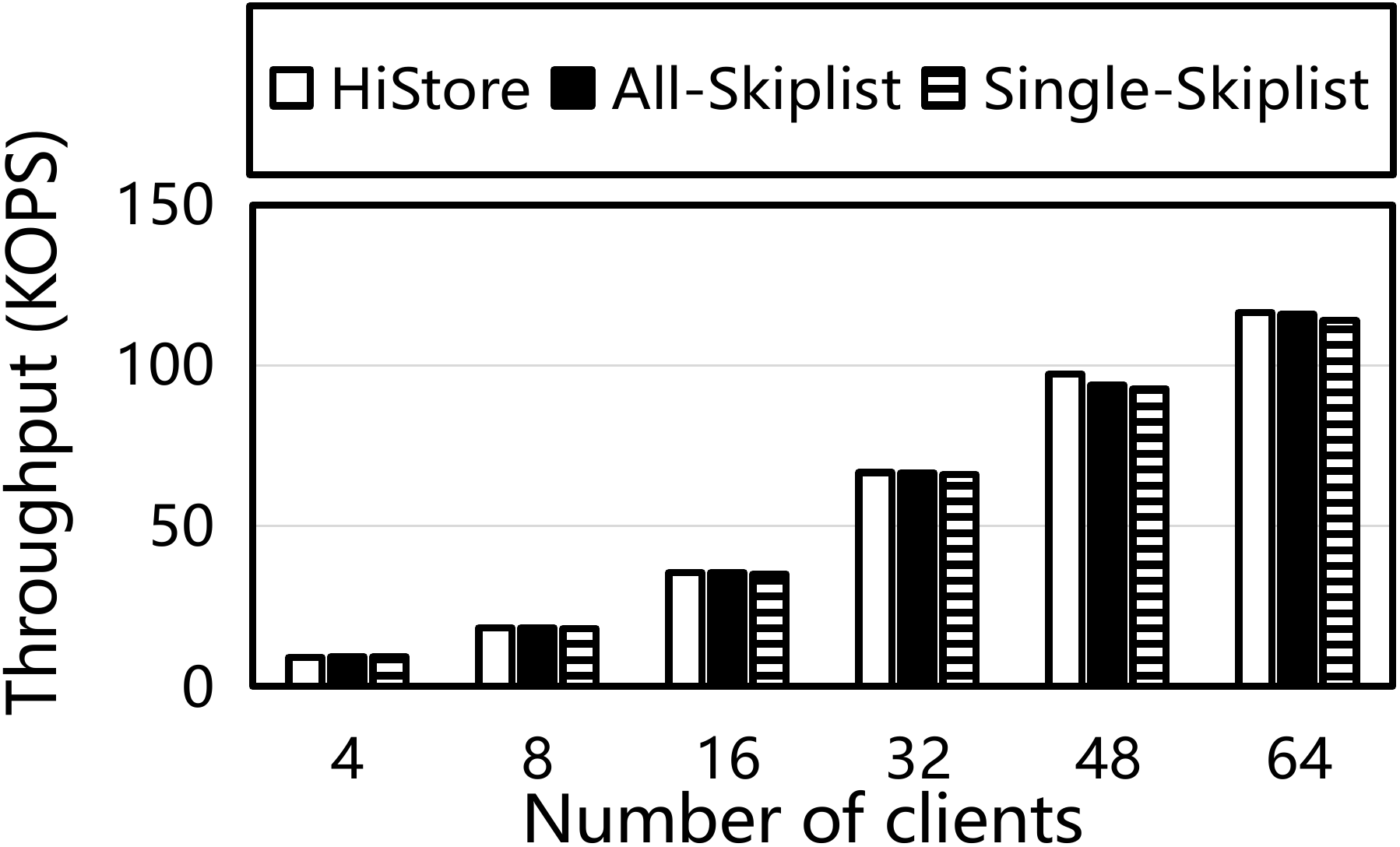}}
	\caption{\textbf{The throughput of basic operations.} \textit{}}
	\label{fig:basic_throughput}
\end{figure*}

\begin{figure*}[ht]
	\centering
	\subfloat[PUT operation]{
		\label{fig:basic_put_lat}
		\includegraphics[width=0.3\textwidth]{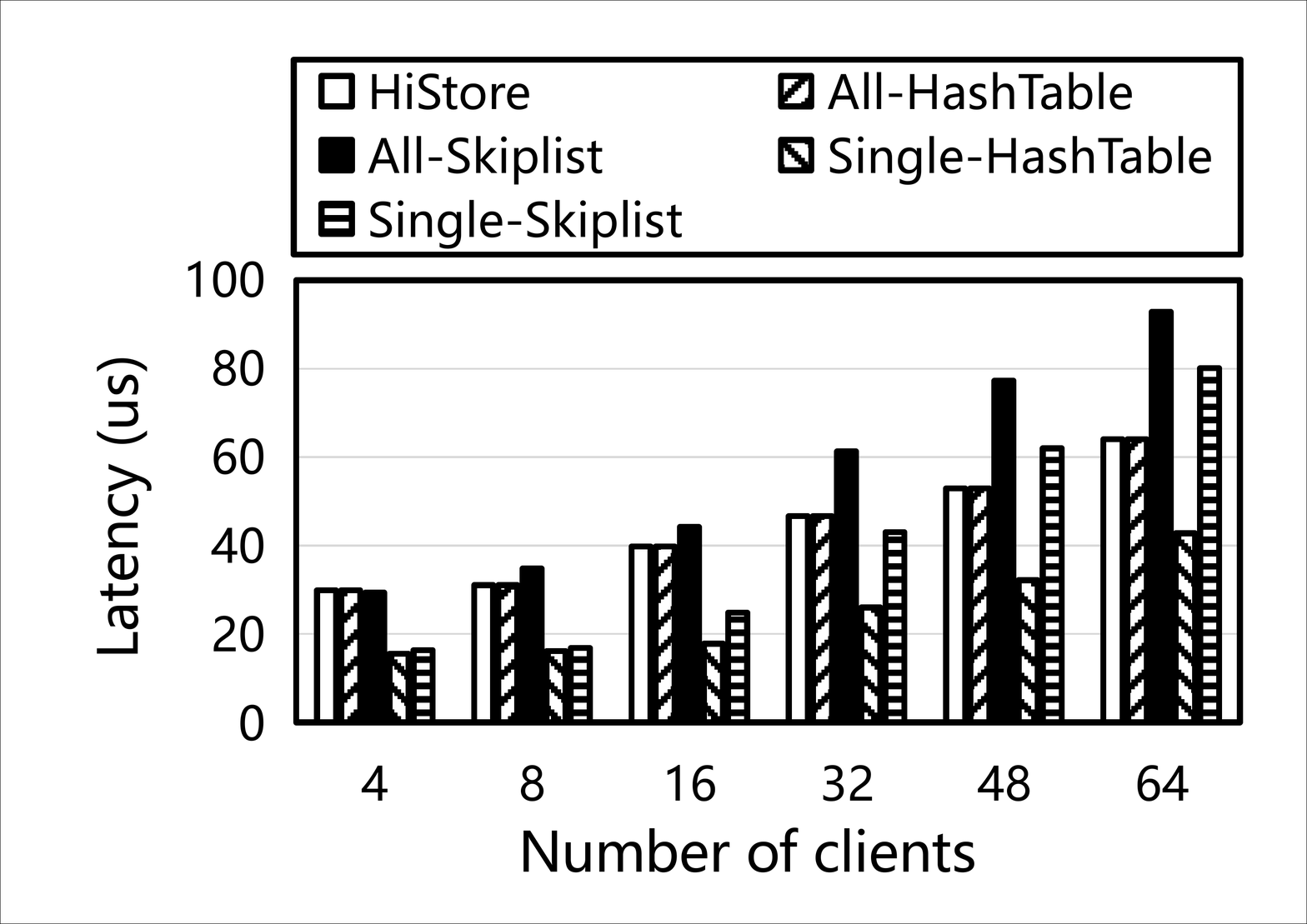}}
	\hspace{10pt}
	\subfloat[GET operation]{
		\label{fig:basic_get_lat}
		\includegraphics[width=0.3\textwidth]{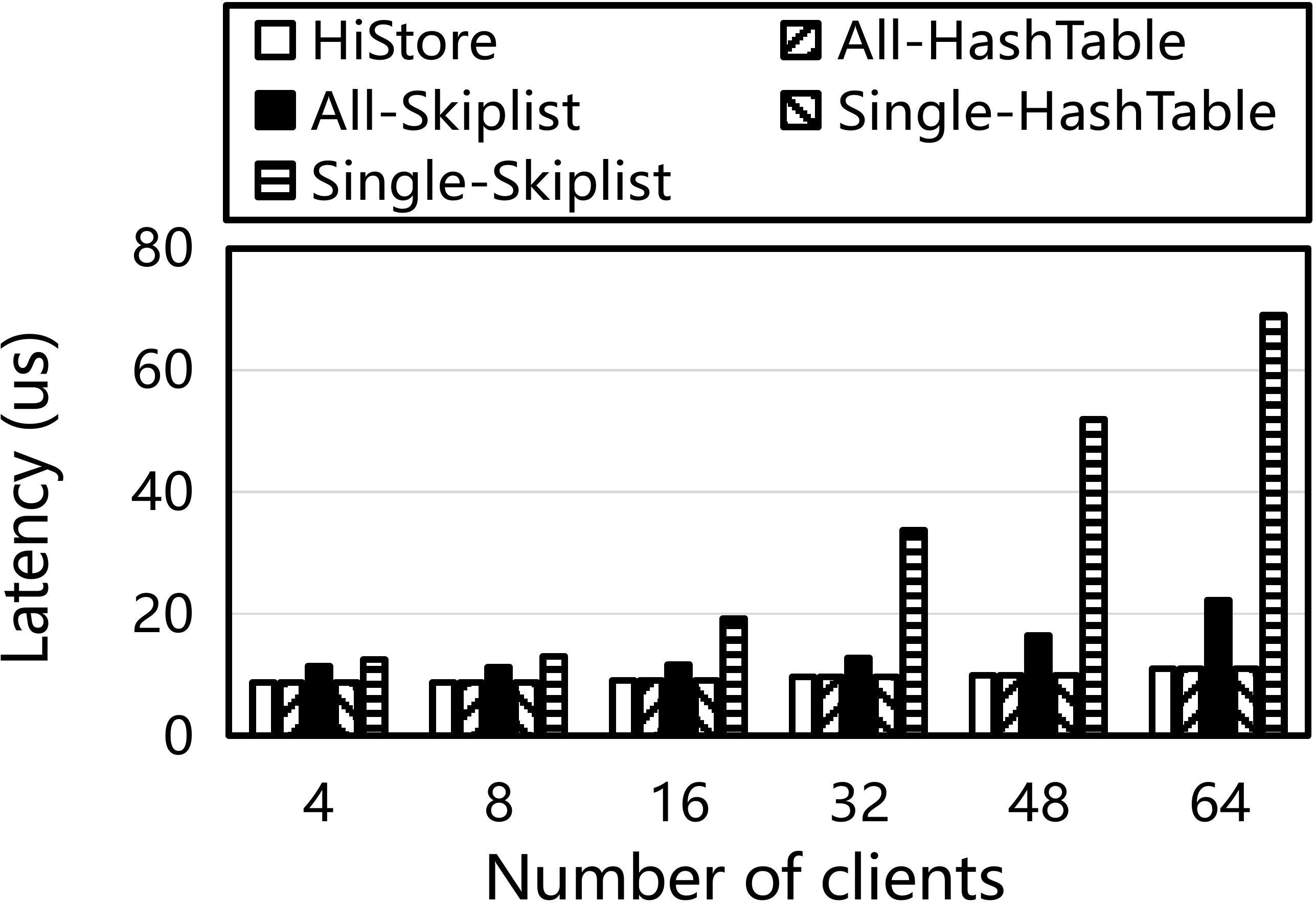}}
	\hspace{10pt}
	\subfloat[SCAN operation]{
		\label{fig:basic_scan_lat}
		\includegraphics[width=0.3\textwidth]{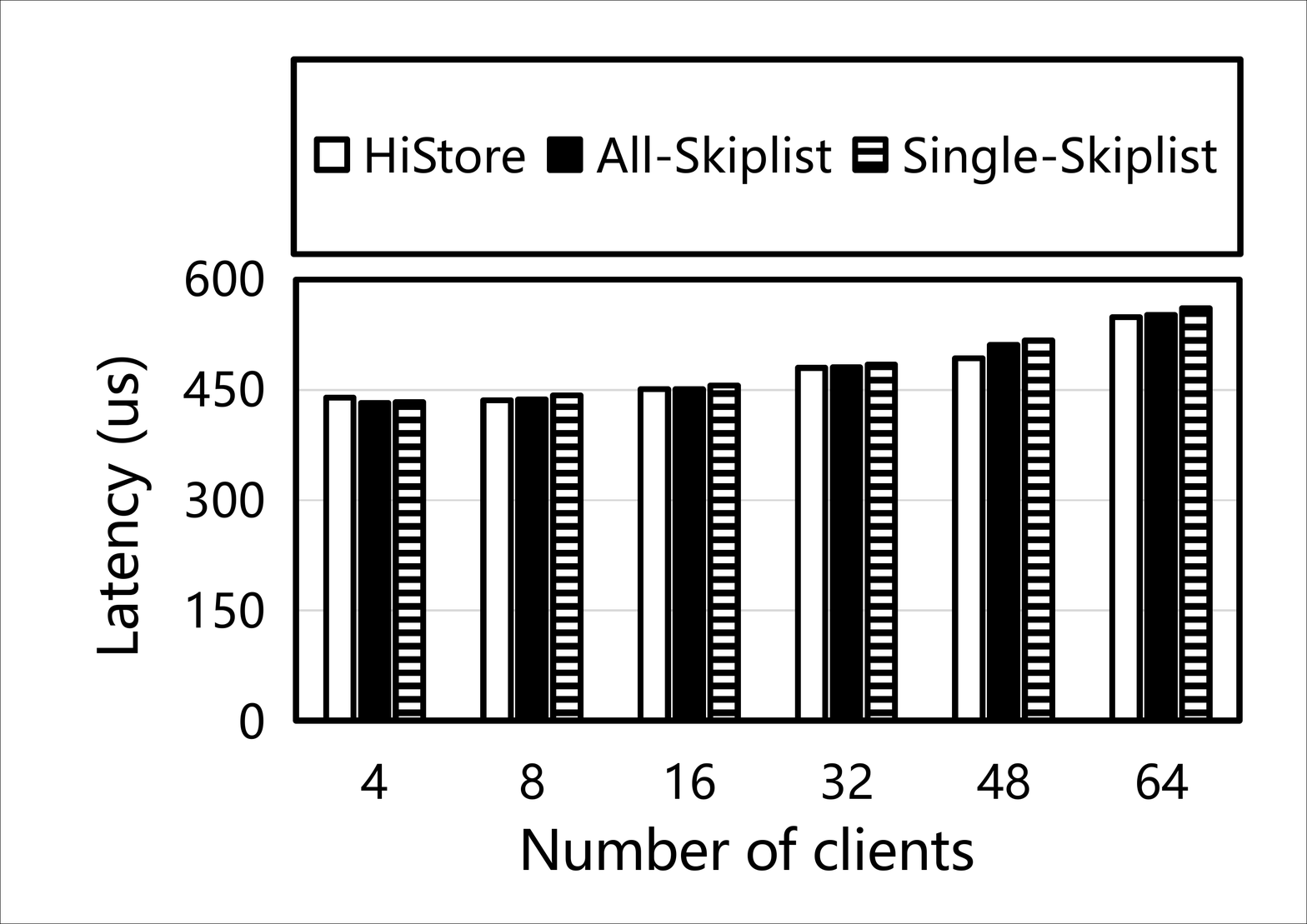}}
	\caption{\textbf{The latency of basic operations.} \textit{}}
	\label{fig:basic_latency}
\end{figure*}

\subsection{Experiment Setup}
\label{sec:setup}

We evaluate the performance of \sysname on a local cluster. Our local cluster consists of five
servers, each of which runs CentOS Linux release 7.6.1810 with 4.18.8 kernel and is equipped with
two Intel Xeon Gold 5215 CPU (2.5\,GHZ), 64\,GB memory and one 100\,Gbps Mellanox ConnectX-5 
Infiniband NIC.
All machines are connected via a 100\,Gbps switch.
We use one machine for data storage, one to send requests as a client, and deploy an index group
consisting of a primary server and two backup servers on the remaining machines.
We use db\_bench~\cite{LevelDB, RocksDB} and YCSB~\cite{YCSB} benchmark to evaluate the performance
of \sysname.
We run five times for each experiment and plot the average result.

We implement another two systems called \textit{all-hashtable} and \textit{all-skiplist} for
comparison, as none existing system realizes hybrid index based on RDMA.
The all-hashtable keeps three hash tables on three servers, while the all-skiplist maintains a
skiplist on each server.
For PUT operations, both of the systems work as \sysname where the primary server sends the index
updates to the backup servers for asynchronous updates, and updates the local index on the primary
server. The all-hashtable allows the client to directly access the hash tables via one-sided verbs,
but does not support range queries.
In the all-skiplist, the client sends GET and SCAN operations to a randomly-chosen server using
two-sides verbs.
Moreover, we compare \sysname with a single server running a hash table (i.e., \textit{single-hashtable}) or
a skiplist (i.e., \textit{single-skiplist}).
For the hash table used, we allocate more buckets to avoid resizing, i.e., the amount of keys
that the hash table can store is higher than the number of keys to store.

\subsection{Performance of Basic Operations}
\label{sec:basic_operation}

To evaluate the performance of basic operations in \sysname, we first load 100 millions (100\,M)
key-value pairs with key size of 16\,B and value size of 32\,B. We then issue 20\,M PUT requests,
20\,M GET requests, and 1\,M SCAN requests using db\_bench.
We start four RPC threads and four worker threads on each index server.
We collect the performance of basic operations under different number of client threads ranging from
4 to 64.
We plot the throughput and latency of PUT, GET, and SCAN operations in
Figure~\ref{fig:basic_throughput} and ~\ref{fig:basic_latency}. 

For the PUT operations, \sysname achieves similar performance to all-hashtable, and higher
performance than all-skiplist.
Compared to single index, the performance of \sysname is lower than that of single-hashtable, and higher
than that of single-skiplist when there are more than 48 client threads.
Figure~\ref{fig:basic_put_tp} and~\ref{fig:basic_put_lat} show the throughput and latency of
PUT operations.
The workflow of processing a PUT operation in \sysname is almost the same as that in all-hashtable and
all-skiplist. The only difference is that \sysname and all-hashtable update the hash table, while
all-skiplist update the skiplist on the primary server.
Thus, the PUT performance of all-skiplist is lower than that of \sysname and all-hashtable
when there are more than four client threads, because updating a skiplist takes longer time than
updating a hash table.
The PUT throughput of \sysname is 1.45 times of that of all-skiplist when there are 64 clients.
The single-hashtable achieves the lowest latency among all schemes. Compared with the
single-hashtable, \sysname increases the latency by 50-124\% because \sysname requires to
send the index updates to other backup servers and wait for their responses.
When the number of clients is small, \sysname achieves lower performance than single-skiplist as
single-skiplist only needs to update one skiplist. The latency of single-skiplist increases
with the number of clients because the server CPU becomes the bottleneck. \sysname reduces the
latency of single-skiplist by 15-20\% when the number of client threads exceeds 48.

\sysname achieves the best read performance among all schemes by leveraging its hybrid index for GET
and SCAN operations.
Figure~\ref{fig:basic_get_tp} and~\ref{fig:basic_get_lat} depict the throughput and latency of GET
operations. 
The GET performance of \sysname is similar to that of single-hashtable and all-hashtable,
because the client can directly access the hash table using one-sided verbs in all three schemes.
\sysname achieves higher performance than both all-skiplist and single-skiplist.
The latency of GET operations in all-skiplist and single-skiplist increases with the number of
the clients, as these two schemes incur CPU cost during key lookups.
Note that the latency of single-skiplist increases significantly while that of all-skiplist
increases slowly, meaning that three skiplists running on three servers relax the CPU burden on one
server.
The GET throughput of \sysname is 2.03 times of that of all-skiplist under 64 clients.
Figure~\ref{fig:basic_scan_tp} and~\ref{fig:basic_scan_lat} show the throughput and latency of SCAN
operations.
\sysname achieves similar SCAN performance to single-skiplist and all-skiplist. The number of
keys covered by each scan operation in our evaluation is 100.   
Although all three schemes use different number of skiplists for range queries, there is nearly no 
performance difference, because the SCAN latency depends on the time of data access rather than the
indexing time (shown in Section~\ref{subsec:eval_micro}).

\subsection{Microbenchmarks}
\label{subsec:eval_micro}

\begin{figure}[t]
	\centering
	\subfloat{
		\includegraphics[width=0.35\textwidth]{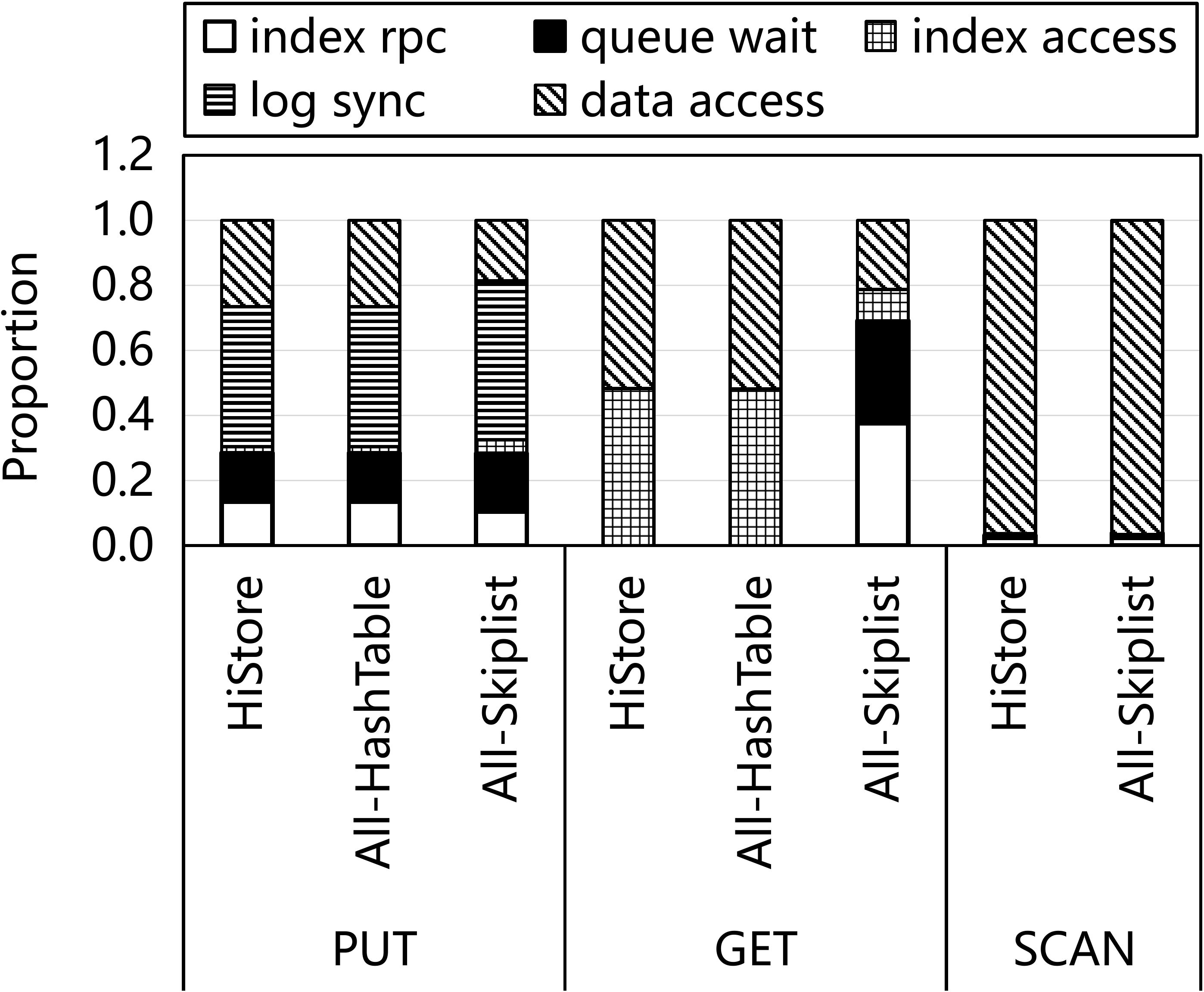}}
	\caption{\textbf{Performance breakdown of basic operations.} \textit{}}
	\label{fig:breakdown}
\end{figure}

We conduct microbenchmarks for \sysname, all-hashtable, and all-skiplist using db\_bench.
We measure the time of the following phases when handling a request:
(i) \textit{index rpc}, the communication time between a client and an index server based on eRPC;
(ii) \textit{queue wait}, waiting in a queue to process;
(iii) \textit{index access}, performing update or search on an index;
(iv) \textit{log sync}, log synchronization between the primary server and backup servers during
writes;
and (vi) \textit{data access}, storing values on data servers or retrieving data based on the value
address.
Figure~\ref{fig:breakdown} shows the performance breakdown of basic operations with 64 clients.
For PUT operations, the time of log sync takes the largest fraction (43-49\%) of the whole latency in all three
schemes. The latency of index access (i.e., inserting to a hash table) in
\sysname and all-hashtable is quite short which can be ignored during writes, while the time of
updating a skiplist takes 4\% of the total write time.
\sysname has the same performance breakdown as all-hashtable for GET operations based on
one-sided verbs, where the index access and data access takes 48\% and 52\% of
the read time respectively.
For all-skiplist, the index rpc and queue wait take 70\% of the read time in total, while data
access only takes 21\%.
When handling a SCAN operation, \sysname and all-skiplist spend about 96\% of the time in data
access while the remaining time is used for index queries.

\subsection{Performance under YCSB Workloads}
\label{sec:ycsb}

\begin{figure}[t]
	\centering
	\subfloat[Normalized throughput]{
		\label{fig:YCSB_tp}
		\includegraphics[width=0.35\textwidth]{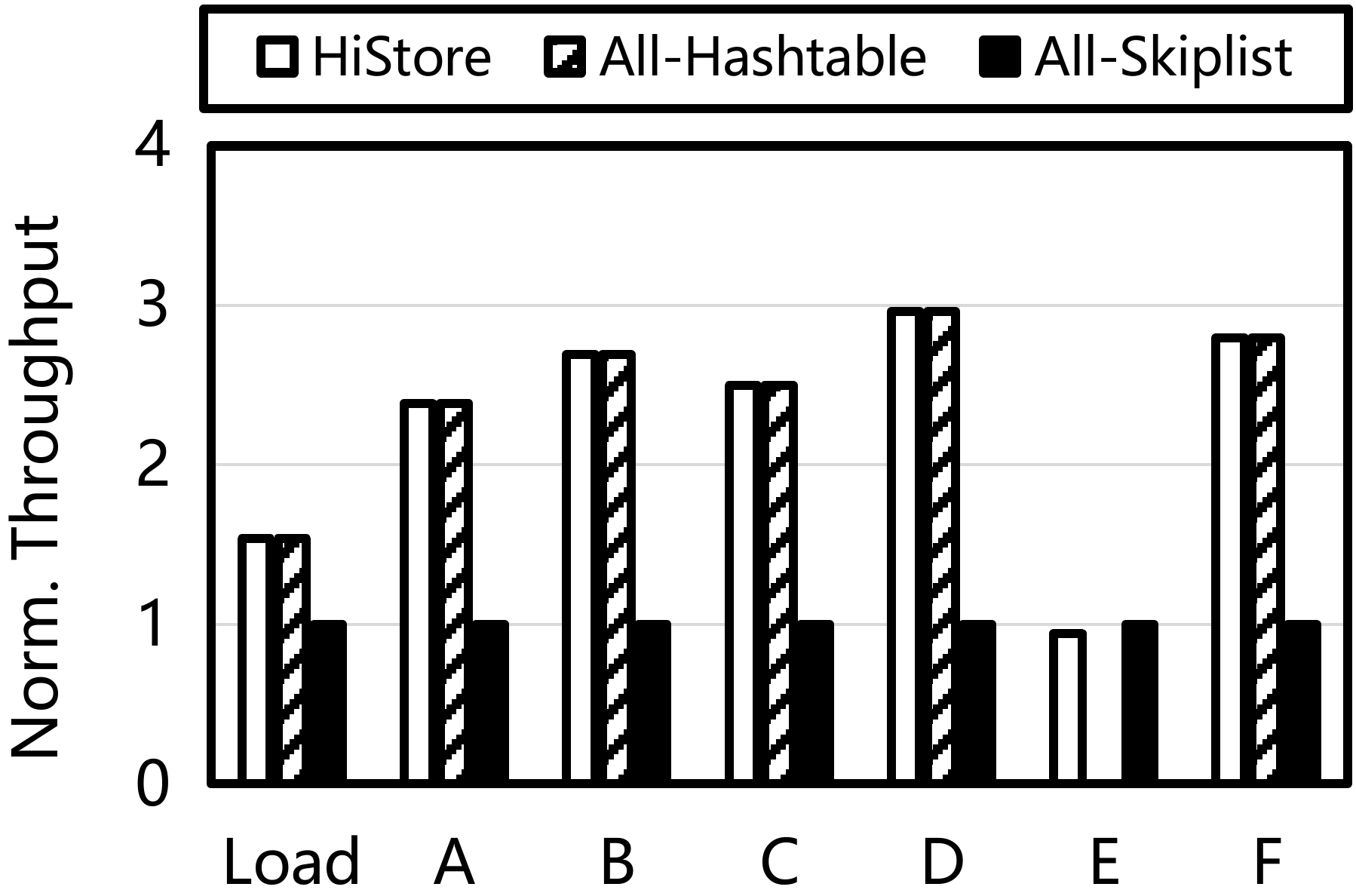}}

	\subfloat[Normalized latency]{
		\label{fig:YCSB_lat}
		\includegraphics[width=0.35\textwidth]{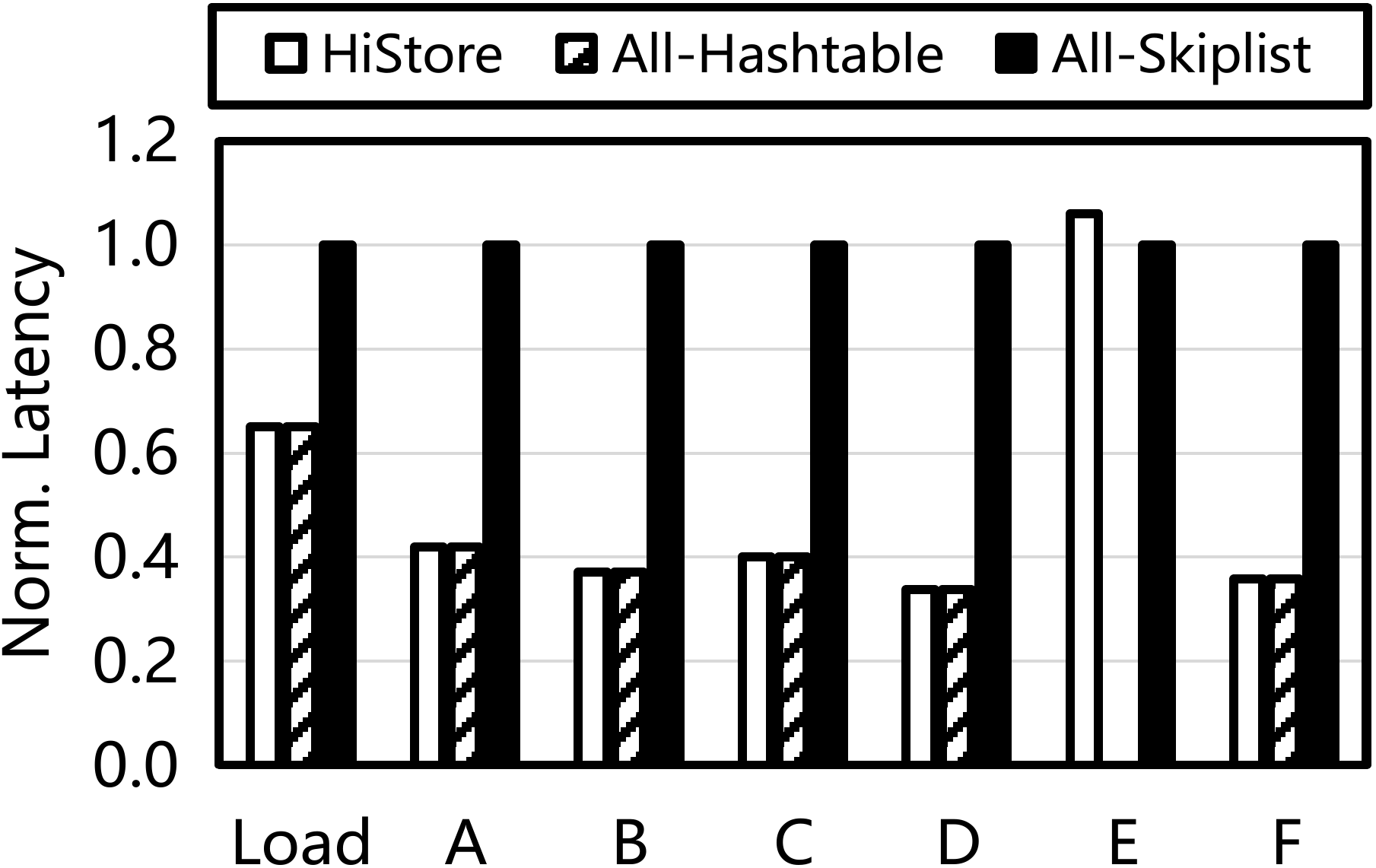}}

    \vspace{-4pt}
	\caption{\textbf{Performance under YCSB workloads.} \textit{This figure shows the throughput and
	latency normalized to the all-skiplist under YCSB workload A (50\% reads and 50\% updates), B (95\% reads
	and 5\% updates), C (100\% reads), D (95\% reads and 5\% inserts), E (95\%
    scan and 5\%
	inserts), and F (50\% reads and 50\% read-modify-writes).}}
	\label{fig:ycsb}
\end{figure}

We evaluate the performance of \sysname under different YCSB workloads. 
We use the default setting with key size of around 20\,B, value size of 32\,B, and a Zipfian request
distribution with a Zipfian constant of 0.9.
The number of keys requested by each scan operation is 100.
We first load 100\,M key-value pairs, and execute each workload by issuing 20\,M requests.
Figure~\ref{fig:YCSB_tp} and~\ref{fig:YCSB_lat} depict the throughput and latency of \sysname under
six YCSB workloads respectively, which are normalized to that of all-skiplist.
The performance of \sysname is close to that of all-hashtable.
Compared to all-skiplist, \sysname increases the throughput by 139-196\% under the workloads without
range queries, because \sysname supports GET operations using the hash table.
For workload E that involves 95\% range queries and 5\% insert, \sysname achieves similar performance
to all-skiplist by leveraging the skiplist in the hybrid index.

\subsection{Recovery and Degraded Performance}
\label{sec:recovery_test}

\begin{figure}[t]
	\centering
	\subfloat{
		\label{fig:recovery_time}
		\includegraphics[width=0.3\textwidth]{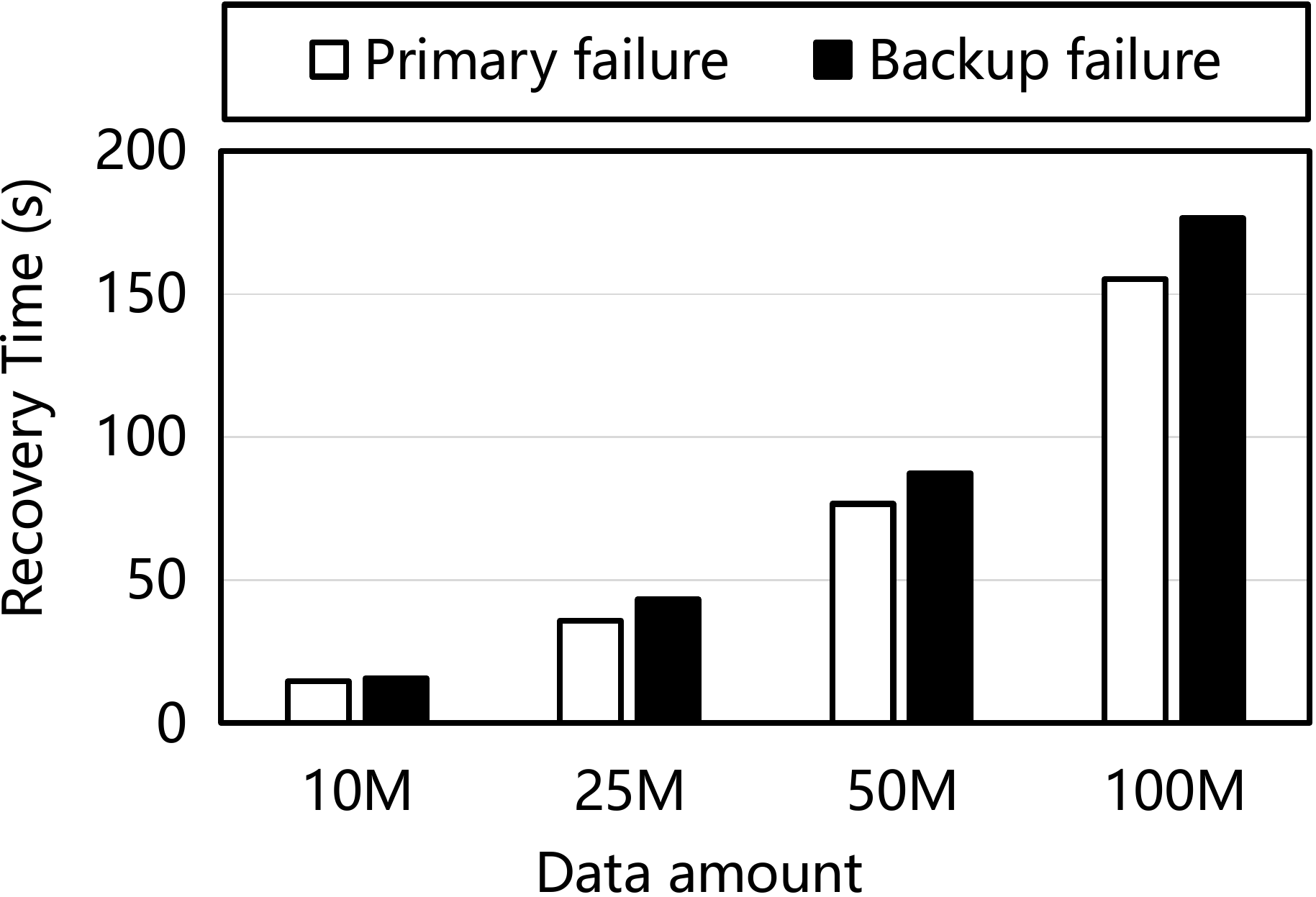}}
    \vspace{-4pt}
	\caption{\textbf{Recovery time in case of the primary and backup server failure.}}
	\label{fig:recovery}
\end{figure}

\begin{figure}[t]
	\centering
	\subfloat[Normalized throughput]{
		\label{fig:recovery_tp}
		\includegraphics[width=0.22\textwidth]{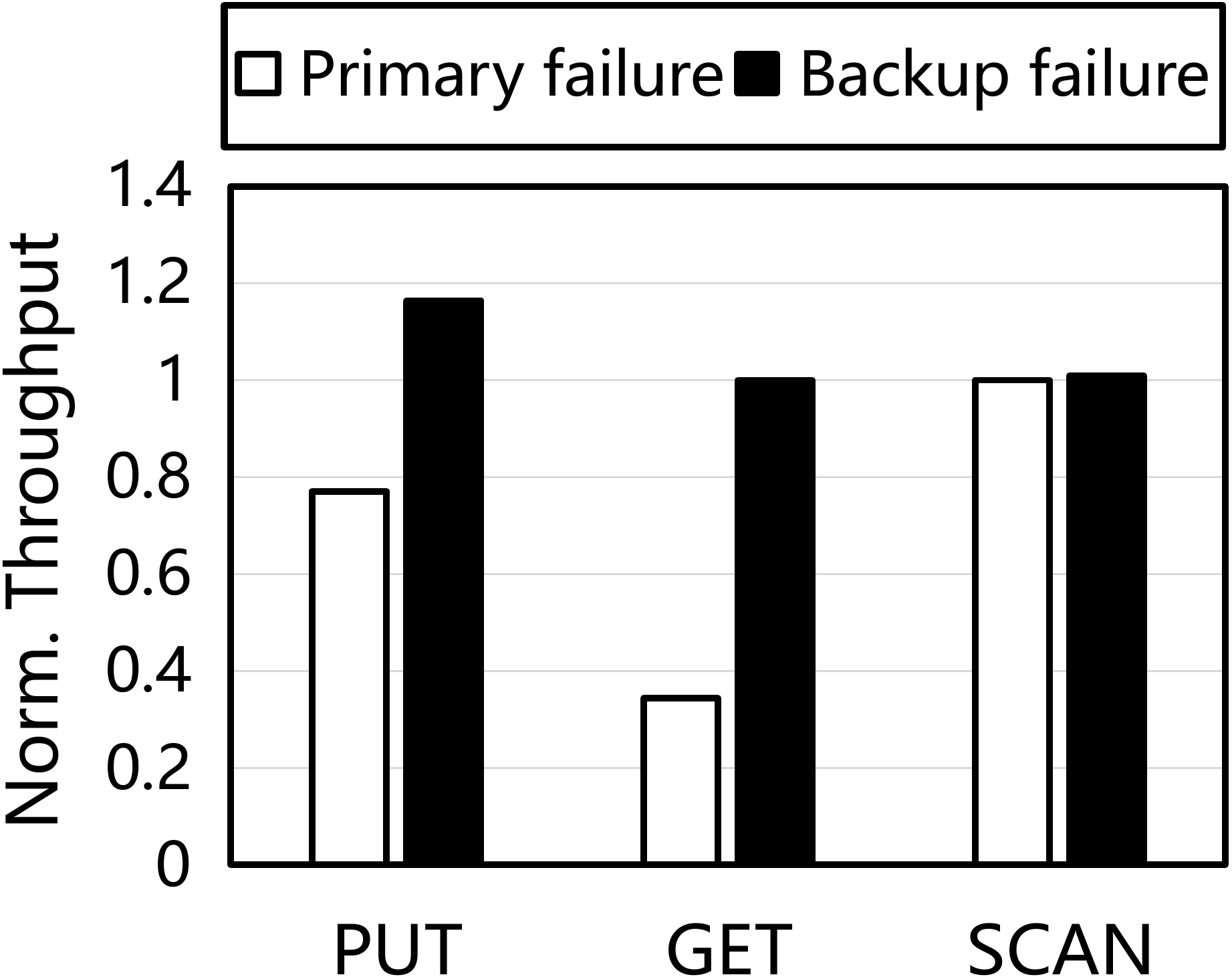}}
	\subfloat[Normalized latency]{
		\label{fig:recovery_lat}
		\includegraphics[width=0.22\textwidth]{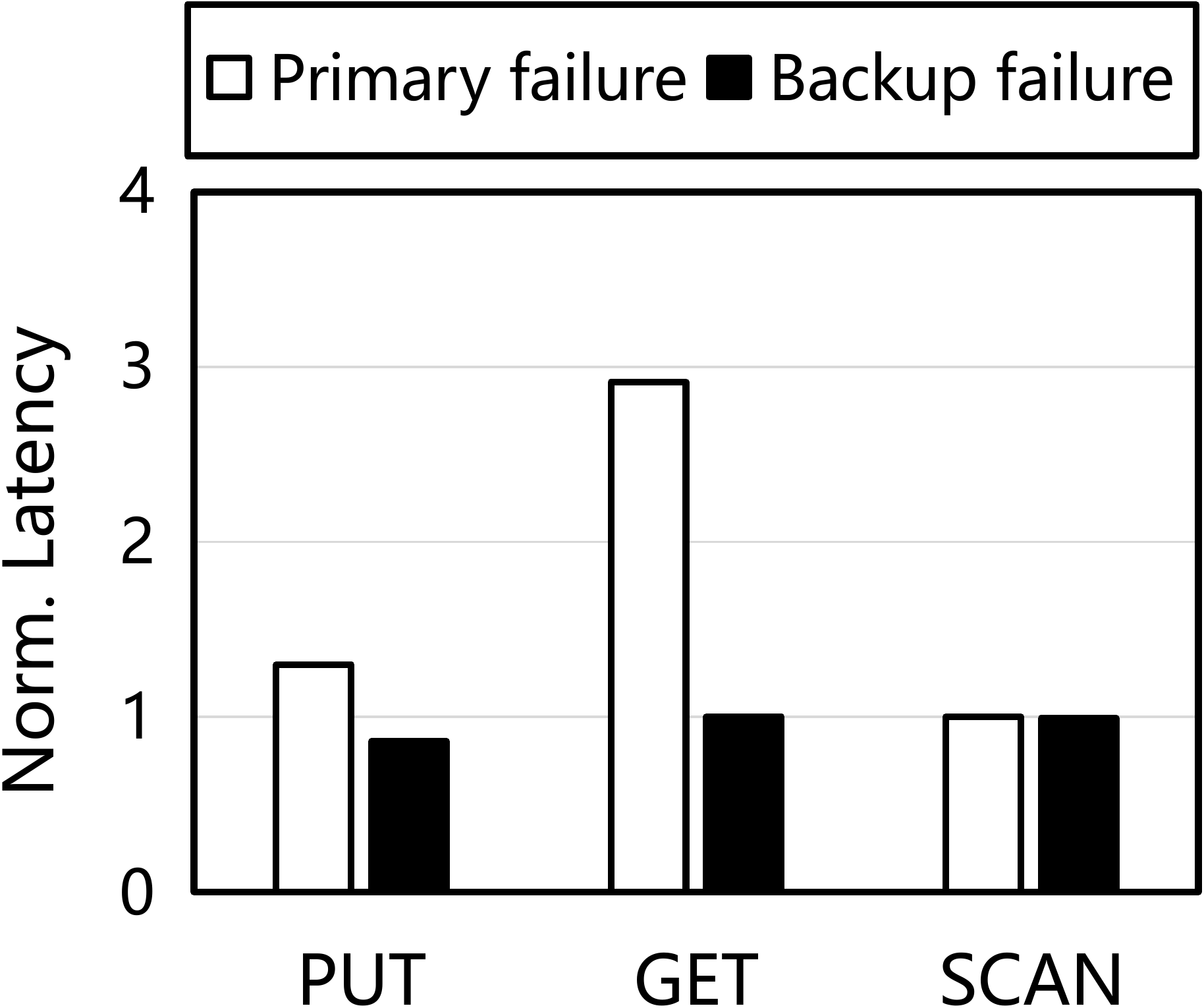}}
    \vspace{-4pt}
	\caption{\textbf{Degraded performance under one index server failure.} \textit{The throughput
	and latency are normalized to the normal performance of \sysname.}}
	\label{fig:degraded_perf}
\end{figure}

We measure the recovery time in case of the primary server and backup server failures, i.e., the
latency of rebuilding a hash table or a skiplist.
Figure~\ref{fig:recovery} shows the recovery time of the primary server and a backup server.
When the number of keys increases from 10\,M to 100\,M, it takes 14-155\,s and 15-176\,s to recover
the primary server and a backup server respectively.
Compared to rebuilding the hash table, the reconstruction of the skiplist is longer it incurs more
CPU cost.

We also evaluate the degraded performance of \sysname when the primary server or one backup server
fails.
Figure~\ref{fig:degraded_perf} illustrate the degraded throughput and latency normalized to the
normal performance of \sysname.
In case of primary server failure, the PUT and GET performance reduces, because \sysname needs to update a
skiplist instead of a hash table and provides single-key lookups based on the skiplist.
When a backup server fails, the PUT performance increases as the primary server only requires to
update one backup server, while the GET performance is not affected.
The performance of range queries does not change in the presence of single-server failures, as there
is at least one index server with skiplist.

%% file: 6_relatedwork.tex
\section{Related Work}
\label{sec:related}

\noindent\textbf{Single index on RDMA.}
A large body of research on RDMA-based key-value store in the literature has explored single index
structure, either hash-based or sorted index. These works are orthogonal to \sysname,
i.e., \sysname can utilize two different types of indexes (i.e., one hashing index and one sorted
index) to combine their benefits.

Many RDMA-enabled key-value stores leverage hashing index to achieve fast lookup services, and
further optimize the usage of hash tables on RDMA~\cite{mitchell13pilaf, dragojevic14farm,
kalia14herd, wang15hydradb, wei15drtm, kalia2016fasst, li2017kv-direct, cassell17nessie, zuo21race}.
Pilaf~\cite{mitchell13pilaf} uses a $n$-way cuckoo hashing algorithm~\cite{pagh04cuckoo} to compute $n$ different hash
buckets for every key by $n$ orthogonal hash functions, where $n = 3$ achieves the best memory
efficiency.
FaRM~\cite{dragojevic14farm} proposes a chained associative hopscotch hashing~\cite{herlihy08hopscotch} to
achieve high space efficiency and a small number of RDMA reads for lookups.
HydraDB~\cite{wang15hydradb} proposes a cache-friendly compact hash table based on the consistent
hashing algorithm~\cite{karger1997}.
DrTM~\cite{wei15drtm} presents cluster hashing which is similar to a chained hashing with
associativity.
RACE~\cite{zuo21race} is a one-sided RDMA-conscious extendible hashing index that supports lock-free
remote concurrency control and efficient remote resizing.

Some key-value stores adopt sorted index with RDMA to support range queries efficiently~\cite{dragojevic15farm,
mitchell16cell, chen16drtmr, kalia19erpc, ziegler19, wei20xstore}.
The updated version of FaRM~\cite{dragojevic15farm} that supports distributed transactions leverages B-Tree
for range queries.
Cell~\cite{mitchell16cell} proposes a hierarchical B-tree where a global tree consists of local trees.
DrTM-R~\cite{chen16drtmr} provides an ordered store in the form of a B$^{+}$-tree in DBX~\cite{wang14dbx}.
Masstree+eRPC~\cite{kalia19erpc} extends Masstree~\cite{masstree12}, an in-memory ordered key-value store,
with eRPC (a RDMA-based RPC library).
Ziegler et al.~\cite{ziegler19} study different design alternatives for tree-based index structure
on RDMA.
XStore~\cite{wei20xstore} maintains a B$^{+}$-tree index at the server, which is implemented by
extending the B$^{+}$-tree index~\cite{wang14dbx} with DrTM+H~\cite{wei18drtmh} (a hybrid RDMA framework).
Sherman~\cite{wang22sherman} is a write-optimized distributed B$^{+}$-tree using local cache, local
and global lock tables based on one-sided verbs, to reduce the number of round trips during writes.
Tebis~\cite{vardoulakis2022tebis} explores to ship the B$^{+}$-tree index on the primary server to
the backup servers over RDMA in LSM-based key-value stores.

\noindent\textbf{Hybrid index in key-value stores.}
Existing works on hybrid index~\cite{xia17hikv, balmau17flodb} mainly consider the usage on one single machine,
while \sysname targets on distributed index based on RDMA.
HiKV~\cite{xia17hikv} proposes a hybrid index consisting of a hashing index~\cite{lim14mica} and B$^{+}$-tree to enable
fast index searching and range queries in DRAM and NVM.
FloDB~\cite{balmau17flodb} presents a hierarchical memory design which is indexed by a small
high-performance concurrent hash table~\cite{david15} and a larger concurrent skiplist~\cite{herlihy12}.
NAM-DB~\cite{zamanian2017nam-db} maps a value of the secondary attribute to a primary key using a
hashing index and a B$^{+}$-tree, but it does not consider the consistency between the two different
indexes and efficient update of the indexes. 

%% file: 7_conclusion.tex
\section{Conclusion}
\label{sec:conclusion}

This paper presents \sysname which leverages hybrid index consisting of a hash
table and a sorted index in RDMA-based key-value store.
\sysname combines the benefit of hashing index and sorted index in a index group, to achieve
efficient single-key lookups and range queries. To minimize the index lookup/update overhead, we
dedicatedly use different RDMA primitives for read/write operations, and apply asynchronous updates
to the sorted index while maintaining strong consistence among different index structures.
Our evaluation results show that \sysname efficiently manages the hybrid index with strong
consistency and provides rich key-value services with low latency and high availability.